\documentclass[twocolumn,superscriptaddress,secnumarabic,amssymb,nobibnotes,aps,prd]{revtex4-2}
\newcommand{\vv}[1]{\mathbf{#1}}
\newcommand{\ii}[0]{\mathrm{i}}

\newcommand{\fran}[1]{\textcolor{black}{#1}}

\usepackage{comment}

\usepackage{natbib}
\usepackage{xcolor}
\usepackage{nicefrac}

\usepackage{hyperref}
\usepackage{here}
\usepackage{amsmath}
\usepackage{bbold}
\usepackage{graphicx}
\begin{document}

\title{Electromagnetic Viscosity in Complex Structured Environments:\\
From black-body to Quantum Friction
}%

\author{M. Oelschl\"ager}%
\affiliation{Humboldt-Universit\"at zu Berlin, Institut für Physik, AG Theoretische Optik \& Photonik, 12489 Berlin, Germany}
\affiliation{dida Datenschmiede GmbH, Hauptstraße 8, 10827 Berlin, Germany}
%\email[REVTeX Support: ]{revtex@aps.org}
\author{D. Reiche}%
\email[Corresponding author: ]{reiche@physik.hu-berlin.de}
\affiliation{Max-Born-Institut, 12489 Berlin, Germany}

% \author{S. Hermann}%
% \affiliation{Humboldt-Universit\"at zu Berlin, Institut für Physik, AG Theoretische Optik \& Photonik, 12489 Berlin, Germany}
\author{C. H. Egerland}%
\affiliation{Humboldt-Universit\"at zu Berlin, Institut für Physik, AG Theoretische Optik \& Photonik, 12489 Berlin, Germany}
\author{K. Busch}%
\affiliation{Humboldt-Universit\"at zu Berlin, Institut für Physik, AG Theoretische Optik \& Photonik, 12489 Berlin, Germany}
\affiliation{Max-Born-Institut, 12489 Berlin, Germany}
\author{F. Intravaia}%
\affiliation{Humboldt-Universit\"at zu Berlin, Institut für Physik, AG Theoretische Optik \& Photonik, 12489 Berlin, Germany}
\date{\today}%
\begin{abstract}
We investigate the nonconservative open-system dynamics of an atom in a generic complex structured electromagnetic environment at temperature $T$.
In such systems, when the atom moves along a translation-invariant axis of the environment, a frictional force acts on the particle.
The effective viscosity due to friction results from the nonequilibrium interaction with the fluctuating (quantum) electromagnetic field, which effectively sets a privileged reference frame.
% We analyze the viscous dynamics with special attention to their temperature dependence.
We study the impact of both quantum or thermal fluctuations on the interaction and highlight how they induce qualitatively different types of viscosity, i.e. quantum and black-body friction.
To this end, we develop a self-consistent non-Markovian description that contains the latter as special cases.
In particular, we show how the interplay between the nonequilibrium dynamics, the quantum and the thermal properties of the radiation, as well as the confinement of light at the vacuum-material interface is responsible for several interesting and intriguing features.
Our analyses is relevant for a future experimental test of noncontact friction and the resulting electromagnetic viscosity.
  \end{abstract}
\maketitle
%
% INTRODUCTION
% ------------
%
\section{Introduction}
%
% Brief introduction into quantum-classical crossover regime
%
Breaking the Lorentz invariance in open quantum systems leads to a number of intriguing effects, regardless of the complexity of the system's configuration~\cite{noether18}.
For example, in the Fulling-Davis-deWitt-Unruh effect \cite{fulling1976,davies75,dewitt75,unruh76} an atom moving in the quantum vacuum perceives its surrounding as a thermal field when it is uniformly accelerated. When its acceleration is not uniform, as in the dynamical Casimir effect, radiation is emitted \cite{dalvit00,dodonov20}. In such situations, an external agent is needed to sustain the motion and to work against a drag force acting on the particle, which tends to restore the inertial dynamics \cite{jaekel93}.
As noted by Einstein and Hopf for a Brownian oscillator \cite{einstein10a,milonni81,ford85}, a drag force also appears
when the atom is moving with respect to a thermal field \cite{mkrtchian2003,lach12,lach12a,jentschura15}. In this case, the nonzero-temperature part of the black-body spectrum sets a preferred inertial frame with respect to which freely moving particles tend to have zero velocity on average.
This drag force, also called black-body friction, has been investigated in different scenarios including constant relativistic \cite{dedkov14a,volokitin15} and
non-relativistic \cite{mkrtchian2003,maianeto04} (see also Refs. \cite{jaekel93,machado02}) velocities. Its impact on the atomic
linewidth-broadening \cite{lach12,lach12a,jentschura15} was also considered in connection to atomic clocks \cite{Kaur20}.

\begin{figure}[t]
  \includegraphics[width=0.6\linewidth]{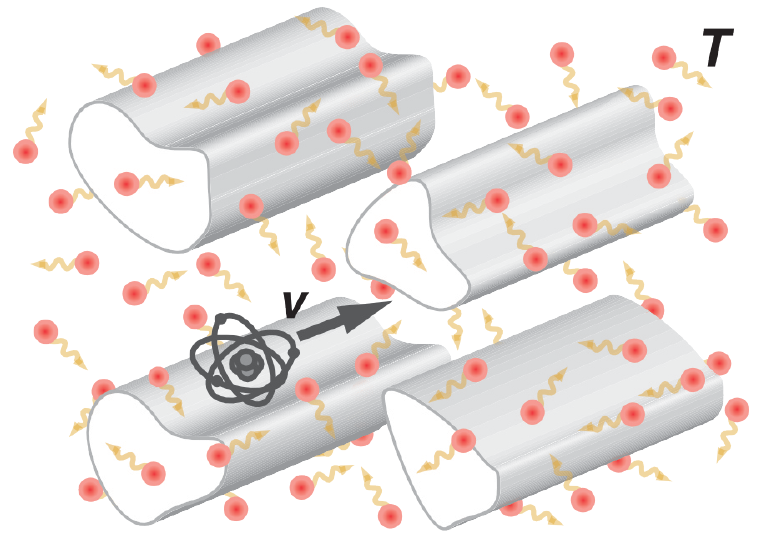}
  \caption{The system under investigation: An atom moving parallel to
an arrangement different, translationally invariant objects surrounded by
thermal radiation at temperature $T$.\label{fig:setup}
}
\end{figure}

Another option to break Lorentz invariance is to introduce an object in the close proximity of the moving particle.
In this case, even at zero temperature and constant velocity, a frictional force arises due to the interaction with the material-modified electromagnetic vacuum field \cite{dedkov2017} which for $T=0$ is usually referred to as quantum friction \cite{pendry1997}.
At finite temperature $T$, constant velocity $v$ and in presence of another object, black-body and quantum friction are the manifestations of a phenomenon that we call electromagnetic viscosity
\begin{align}
\label{Eq:ViscosDef}
\mu
%(v,T)
  &=
  F/v,
\end{align}
where $F$ the is total electromagnetic drag acting on the atom.
Although the viscosity has already been investigated in various contexts \cite{nourtier77,tomassone1997,pendry1997,hoye2015a,volokitin2007,jentschura15,intravaia11a,dedkov2017}  -- including decoherence \cite{viotti19}, thermodynamic considerations \cite{guo21,hsiang21}, its connection to Cherenkov \cite{volokitin16b} and Hawking radiation \cite{intravaia2016b,reiche21} -- some interesting and relevant features have been overlooked.
In the following, we generalize the earlier findings by incorporating the net transfer of angular momentum from the field to the particle. To this end, we go beyond common approximations and employ a fully self-consistent description of the system. The transfer of angular momentum arises from the so-called spin-momentum locking of light \cite{oinakamura1981,bliokh2015a}, a behavior induced by the natural confinement of the electromagnetic field in the vicinity of an interface of two different materials \cite{intravaia2019}.
In particular, since this effect has been already considered for quantum friction \cite{intravaia2019}, we focus here on the
temperature dependence of the viscosity, bridging the gap towards black-body friction. We show that the thermal part of the electromagnetic viscosity naturally decomposes into three contributions
\begin{align}
\mu
%(v,T)
  \to
  \mu_T
  =
  \mu_T^{\mathrm{vac}}+ \mu_T^t + \mu_T^r,
\end{align}
where (i) $\mu_T^{\mathrm{vac}}$ describes the interaction with the thermal quantum vacuum field,
(ii) $\mu_T^t$ includes the interaction between the material-modified field and the atomic translational degrees of freedom, and (iii) $\mu_T^r$ corresponds to the contribution involving the exchange of angular momentum.
Especially the latter term requires the inclusion of the full backaction from the environment onto the dynamics of the particle and, as we will see, tends to reduce the drag experienced by the particle.
It is not covered by traditional approximate or equilibrium-based techniques, and -- to the best of our knowledge -- has not
been analyzed in the literature yet.

The manuscript is structured as follows.
In the next Section \ref{sec:noncontact} we introduce our model, derive the main results of the present manuscript and compare our formulas to the existing literature.
In Section \ref{sec:quacacross}, we discuss the viscosity [see Eq.~\eqref{Eq:ViscosDef}] in various contexts, specify our result to the situations of an atom moving in close vicinity to a dissipative plate and provide both analytical asymptotes as well as numerical evaluations.
In Section \ref{sec:noneqflu}, we analyze our findings from a different perspective by exploring the spectral density of the interaction between atom and field.
We close our discussion with summarizing remarks in Section \ref{sec:conclusion}.

%
% NONCONTACT FRICTION
% -------------------
%
\section{Dipole Force on a moving atom}
\label{sec:noncontact}
We consider an electrically neutral atom that is propelled by an external agent along the $x$-direction, corresponding to the direction of translational invariance of a complex macroscopic electromagnetic environment -- i.e. $N$ translationally invariant objects. For simplicity, we neglect magnetic properties of both the atom and its surroundings.
The atomic trajectory is described by $\mathbf{r}_a(t)=x_{a}(t) \mathbf{x}+\mathbf{R}_a$  (see Fig.~\ref{fig:setup}), where $x_{a}(t)$ gives the details of the atom's motion and $\mathbf{R}_a\equiv(y_{a},z_{a})$ denotes the transversal coordinates, i.e. the position of the atom in the lateral plane.
We consider geometric dimensions such that a dipole description of the particle is sufficient and we can neglect higher order multipoles.
The atom is then described by its electric dipole operator $\hat{\mathbf{d}}$ and the total electric field by the operator $\hat{\mathbf{E}}$.
For simplicity, we also assume the total system's density matrix at the initial time $t_0$ in the far past of the experiment factorizes into $\hat{\rho}(t_0)=\hat{\rho}_\mathrm{\mathrm{atom}}(t_0)\otimes\hat{\rho}_\mathrm{\mathrm{field}}(t_0)$.
At late times, however, due to its dissipative nature, the total system reaches a unique highly entangled nonequilibrium steady-state (see \cite{reiche20a} and references therein), which does not conserve any memory of its initial condition.
In our case the electric field which would exist without the atom, $\hat{\mathbf{E}}_0$, is supposed to be thermalized at temperature $T$ and, therefore, obeys the fluctuation-dissipation theorem \cite{kubo1966}.
In momentum- and frequency-space we can write
\begin{multline}
\label{eq:flucdisE0}
\langle
\hat{\mathbf{E}}_0(q,\mathbf{R}_a,\omega)
\hat{\mathbf{E}}^{\intercal}_0(\tilde{q},\mathbf{R}_a,\tilde{\omega})
\rangle
\\
= 8\pi^2\hbar
\left[1+n(\omega)\right]
\underline{G}_\Im(q,\mathbf{R}_a,\omega)
\delta(\omega+\tilde{\omega})\delta(q+\tilde{q})
\,,
\end{multline}
where $q$ is the component of the wave vector along the $x$-direction, $\omega$ the frequency of the radiation and the superscript ``$\intercal$'' gives the transpose of an expression rendering the previous product of operators as a dyadic (see appendix \ref{App:DerForceGen}).
$\underline{G}$ is Green tensor that solves Maxwell's equations with appropriate boundary conditions for our setup. Since in the following we will be interested in positions having the same transversal coordinates, $\mathbf{R}_a$ will appear only once in the argument of the Green tensor.
The angular brackets denote the quantum average over the initial state density matrix $\hat{\rho}(t_0)$, the subscript ``$\Im$'' indicates $\underline{G}_\Im=(\underline{G}-\underline{G}^\dagger)/(2\ii)$, $\hbar$ is Planck's reduced constant and $\delta(\cdot)$ is the Dirac function.
Also, we defined the Bose occupation number $n(\omega)=(e^{\beta\hbar\omega}-1)^{-1}$ with the inverse temperature $\beta=1/(k_\mathrm{B}T)$ ($k_{\mathrm{B}}$ denotes Boltzmann's constant).

Some general comments can be made about the tensor $\underline{G}_{\Im}$ (see also \cite{dennis03,reiche20a}): For example, in vacuum, due to the transversality of the field, $\underline{G}_{\Im}$ is a symmetric tensor and is even in $q$.
However, in the vicinity of an interface, due to the spatial confinement characterizing the near-field, the radiation can feature a longitudinal component with respect to the direction of motion.
This translates into (off-diagonal) skew-symmetric components in $\underline{G}_{\Im}$ that are odd in $q$.
Mathematically, this encodes the so-called spin-momentum-locking of light \cite{bliokh2015}. As we will see below, the latter has specific consequences in our particular context (see discussion after Eq.~\eqref{Eq:ForceDipoleMat2}).

In the non-relativistic case, the motion-induced drag experienced by the atom is then given by the $x$-component of the Lorentz force \cite{intravaia2014,intravaia2019}, i.e.
\begin{equation}
\label{Eq:ForceDipole}
  F(t)=
  \lim_{\vv r \to \vv r_a(t)}
  \langle \hat{\vv d}(t)\cdot
  \partial_x
  \hat{\vv{E}}(\vv r, t)
 \rangle
  \,.
\end{equation}
For our purposes it is convenient to write the electric field operator as a sum of $\hat{\vv E}_0$, the uncoupled field introduced above, and $\hat{\vv E}_\mathrm{ind}$ which describes the radiation induced by the atom.
Proceeding as in Ref. \cite{intravaia2016}, we can write
\begin{align}
\label{eq:efield}
\hat{\mathbf{E}}_\mathrm{ind}(\mathbf{r},t)
&=
\int\limits_{t_0}^t \mathrm{d}t'\,
\underline{G}(\mathbf{r},\mathbf{r}_a(t'),t-t')\,
\hat{\mathbf{d}}(t')
\\\nonumber
&=
2\ii
\int\limits_{0}^{t-t_0} \mathrm{d}\tau
\int\frac{\mathrm{d}\omega}{2\pi}
\int\frac{\mathrm{d}q}{2\pi}~
e^{-\ii \omega \tau}
e^{\ii q[x-x_{a}(t-\tau)]}
\\\nonumber
&\qquad\times
\underline{G}_{\Im}(q,\mathbf{R}_a,\omega)\,
\hat{\mathbf{d}}(t-\tau).
\end{align}
For the atomic dipole operator, we focus for simplicity on a dynamics associated with a single transition frequency $\omega_a$ and described in terms of a 3D isotropic oscillator \cite{intravaia2014}
\begin{equation}
  \label{eq:deom}
  \left(\partial_t^2+\omega_a^2\right)
  \hat{\mathbf{d}}(t)
  =
  \alpha_0 \omega_a^2
  \hat{\mathbf{E}}(\mathbf{r}_a(t),t).
\end{equation}
Here, $\alpha_0$ is the static atomic polarizability, describing the coupling between the atom and the electromagnetic radiation.
The details of our model for field and atom have been discussed elsewhere \cite{intravaia2014,intravaia2016,intravaia2019}.

At late times, when the external force and the drag balance out, the atom prescribes approximately the trajectory $x_{a}(t)\sim v t$ with $v$ the constant non-relativistic velocity characterizing the steady-state (see e.g. \cite{reiche20a} and references therein).
We can then solve the system self-consistently up to all orders in the coupling (connected to $\alpha_0$ \cite{reiche20a}) and evaluate the atomic power spectrum  $\underline{S}_v(\omega)$ from the dipole correlator
$
4\pi^2 \delta(\omega+\tilde{\omega}) \underline{S}_v(\omega)
=
\langle \hat{\vv d}(\omega)\hat{\vv d}^\intercal(\tilde{\omega})\rangle
$ \cite{intravaia2014,intravaia2016,intravaia2019,reiche20a}. We obtain
\begin{align}
\label{Eq:FDTNEq}
%4\pi^2
\underline{S}_v(\omega)
  &=
  \underline{\alpha}_v(\omega)
  \underline{\kappa}_v(\omega)
  \underline{\alpha}_v^{\dagger}(\omega)
,
\end{align}
where the extra subscript ``$v$'' explicitly denotes the velocity-dependence of the expressions. The tensors $ \underline{\kappa}_v$ and $\underline{\alpha}_v$ are, respectively, the steady-state nonequilibrium (uncoupled) electric field's power spectrum [Eq.~\eqref{eq:flucdisE0}]
and the steady-state dressed atomic polarizability.
They can be written as
\begin{subequations}
\label{Eqs:SpecPol}
\begin{align}
\label{Eq:FieldSpectrum}
  \underline{\kappa}_v(\omega)
  &=
%  4\pi
  \frac{\hbar}{\pi}
  \int \frac{\mathrm{d}q}{2\pi}
  \left[
  n(\omega_q^+) +1 \right]
  \underline{G}_\Im(q,\mathbf{R}_a,\omega_q^+),
  \\
  \label{Eq:PolDressed}
  \underline{\alpha}_v(\omega)
  &=
  \left[
  \mathbb{1}
  -
  \alpha_{\mathrm{B}}(\omega)
  \int\frac{\mathrm{d} q}{2\pi}~
  \underline{G}(q,\mathbf{R}_a,\omega_q^+)
  \right]^{-1}\alpha_{\mathrm{B}}(\omega)
  ,
\end{align}
\end{subequations}
where $\omega_q^{\pm}=\omega\pm qv$ is the Doppler-shifted frequency, $\mathbb{1}$ is the three-dimensional unit matrix, and
$\alpha_{\mathrm{B}}(\omega)=\alpha_0\omega_a^2/  \left( \omega_a^2 - [\omega+\ii 0^{+}]^2 \right)$ is the (causal) bare polarizability of the isolated atom \cite{intravaia2012a}.
The appearance of the Green tensor in Eq.~\eqref{Eq:PolDressed} indicates that the polarizability is dressed via the interaction with the electromagnetic environment. In particular, it is interesting to note that the coupling between system and environment survives also at zero velocity since $\lim_{v\to 0}\underline{\alpha}_v\neq\alpha_{\mathrm{B}}(\omega)\mathbb{1}$ \cite{maianeto04,intravaia2011}.

Equation \eqref{Eq:FDTNEq} is the finite-temperature generalization of the nonequilibrium fluctuation-dissipation theorem reported, e.g., in Refs. \cite{intravaia2016,intravaia2019}.
It reveals that the particle's and the (field) environment's fluctuations are closely intertwined, highlighting the importance of a self-consistent approach that fully includes backaction from the environment onto the particle \cite{maianeto04,reiche20a}.
More precisely, both the dressed polarizability \emph{and} the field spectrum in Eq.~\eqref{Eq:FieldSpectrum} are in part related to the local field density of states $\rho_{\mathrm{LDOS}}\propto\mathrm{Im}\mathrm{Tr}[\underline{G}]$ \cite{joulain2003}.
Interestingly, however, the complex tensorial structure of the nonequilibrium fluctuation-dissipation theorem in Eq.~\eqref{Eq:FDTNEq} also causes off-diagonal elements of the Green tensor to contribute to the interaction.
In combination with the integral over Doppler-shifted frequencies [see Eqs. \eqref{Eqs:SpecPol}] and spin-momentum locking of light in the vicinity of a surface \cite{lodahl17}, the contribution of these off-diagonal elements (related to the spin local density of states introduced in \cite{intravaia2019,reiche21})
has been reported to involve the atomic rotational degrees of freedom into the frictional interaction, generating a corresponding component for the force \cite{intravaia2019}.
This will become an important part in our discussion of $\mu_T$ [see Eq.~\eqref{Eq:muTtmuTr}].

Lastly, we would like to mention that, in spite of its formal resemblance to equilibrium fluctuation-dissipation theorems \cite{kubo1966}, Eq.~\eqref{Eq:FDTNEq} includes effects that cannot be captured by equilibrium-based techniques.
For comparison, in certain situations (see, e.g., Ref. \cite{reiche20a} and references therein for details) it is viable to consider the \emph{equilibrium} fluctuation-dissipation theorem also for steady-state nonequilibrium systems. Formally this can be motivated by defining spatially separated subsystems which might equilibrate separately, i.e. in our case the atom and the field \cite{dedkov2017}.
In the following we will use this local thermal equilibrium (LTE) approximation as a reference to highlight the difference introduced by our nonequilibrium description.

Returning to the expression for the Lorentz force in Eq.~\eqref{Eq:ForceDipole}, we can write
\begin{align}
\label{Eq:FRewritten}
F(t)=
\lim_{\vv r \to \vv r_a(t)}
2\mathrm{Re}\;\langle\hat{\vv d}(t)\cdot
\partial_x
\hat{\vv{E}}^{\oplus}(\vv r, t)
\rangle,
\end{align}
where $\hat{\vv{E}}^{\oplus}$ is the positive-frequency component of the electric field operator (we refer to Ref. \cite{intravaia2016a} for details).
Notice, that while in Eq.~\eqref{Eq:ForceDipole} the order of the operators is irrelevant, this is not true anymore for the above rewritten version, Eq. \eqref{Eq:FRewritten}.
As a consequence, a specific order has to be preserved in the calculation and, for convenience, we choose normal ordering.
It is well known that, despite that the final result is independent from this choice, the interpretation of the different contributions appearing in the final expression and their attribution to the dipole or the field dynamics might depend on it \cite{Milonni73,Dalibard82,intravaia2016a}.
Inserting our model for the dipole's dynamics [Eqs. \eqref{eq:efield} - \eqref{Eqs:SpecPol}] yields in the steady-state ($-t_{0},t\to\infty$) the constant force (see Appendix \ref{App:DerForceGen})
\begin{align}
  \label{Eq:ForceDipoleLate}
  F
  &=
  -
  2
  \int_0^{\infty}\mathrm{d}\omega
  \int\frac{\mathrm{d}q}{2\pi}
  q
  \\\nonumber
  &~\times
  \mathrm{Tr}
  \left[
  \left\{
  \frac{\hbar}{\pi}
  n(\omega)
  \underline{\alpha}_{v,\Im}(-\omega_q^-)
  +
  \underline{S}_{v}(-\omega_q^-)
  \right\}
  \underline{G}_{\Im}^{\intercal}(q,\mathbf{R}_a,\omega)
  \right],
\end{align}
where we have defined $\underline{\alpha}_{v,\Im}$ analogously to $\underline{G}_{\Im}$ and we have used that
$\underline{G}_{\Im}(q,\mathbf{R}_a,\omega)=-\underline{G}_{\Im}^{\intercal}(-q,\mathbf{R}_a,-\omega)$ as well as the fact that the product of a symmetric and a skew-symmetric matrix vanishes under the trace.
In our approach, the first and the second term in the curly brackets, respectively, correspond to the dipole interacting with the unperturbed field $\hat{\mathbf{E}}_0$ and the induced field $\hat{\mathbf{E}}_{\mathrm{ind}}$.
We note that both terms depend on the temperature [the power spectrum implicitly via the nonequilibrium fluctuation-dissipation theorem in Eq.~\eqref{Eq:FDTNEq}] and that our result recovers the quantum frictional force ($T\to 0$) reported in previous works \cite{intravaia2014,intravaia2016a}.
The latter can be directly seen from $n(\omega)\to 0$ for $T\to0$ ($\omega\ge0$).
The physics of the system imprints some mathematical properties on the quantities appearing in Eq.~\eqref{Eq:ForceDipoleLate}. These can be used to rewrite the above expression in a different but equivalent form.
Indeed, replacing $\omega\to\omega+qv$ and realizing that the integral kernel is an odd function on the interval $q\in(-\infty,\infty)$ and $\omega\in[-qv,0]$ we can rewrite Eq.~\eqref{Eq:ForceDipoleLate} as
\begin{align}
\label{Eq:ForceDipoleLate2}
&F
=
2
\int_{0}^{\infty}\mathrm{d}\omega
\int\frac{\mathrm{d}q}{2\pi}
q
\\\nonumber
&\times
\mathrm{Tr}
\left[
\left\{
\frac{\hbar}{\pi}
[1+n(\omega_q^+)]
\underline{\alpha}_{v,\Im}(\omega)
-
\underline{S}_v(\omega)
\right\}
\underline{G}_{\Im}(q,\mathbf{R}_a,\omega_q^+)
\right].
\end{align}
Here, we have also utilized that $n(-\omega)=-[1+n(\omega)]$, $\underline{\alpha}_{v,\Im}(-\omega)=-\underline{\alpha}_{v,\Im}^{\intercal}(\omega)$
and that the power spectrum fulfills the identity $\underline{S}_v(-\omega)=\underline{S}_v^{\intercal}(\omega)-(\hbar/\pi)\underline{\alpha}_{v,\Im}^{\intercal}(\omega)$ (see Appendix \ref{App:DerForceGen} for details).
The main difference between Eqs.~\eqref{Eq:ForceDipoleLate} and \eqref{Eq:ForceDipoleLate2} lies in a rearrangement of the contributions due to the field's and to the dipole's dynamics. In particular, Eq. \eqref{Eq:ForceDipoleLate2} corresponds to an approach where the splitting related to a positive and negative frequency integration is not performed (see Appendix \ref{App:DerForceGen}).
As expected, the corresponding ordering of operators, when carried out consistently, has no consequences for the observable force.
Importantly, however, the calculation leading to Eqs.~\eqref{Eq:ForceDipoleLate} and \eqref{Eq:ForceDipoleLate2} requires the assumption of a linear relation between the dipole's  and the electric field's dynamics. This is, for example, the case for the isotropic oscillator described in Eq.~\eqref{eq:deom}.
Conversely, the procedure leading to the zero-temperature version of Eq.~\eqref{Eq:ForceDipoleLate} does not depend on the concrete model that describes the atomic internal degrees of freedom \cite{intravaia2014,intravaia2016a}.

It is instructive to check the consistency of our result with related work.
To this end, we note that $\underline{\alpha}_{v,\Im}$ is connected to the Green tensor via the identity
\begin{align}
\label{Eq:FDTSecondKind}
\underline{\alpha}_{v,\Im}(\omega)
  &=
  \underline{\alpha}_v(\omega)
  \int\frac{\mathrm{d}q}{2\pi}~
  \underline{G}_{\Im}(q,\mathbf{R}_a,\omega_q^+)
  \underline{\alpha}^{\dagger}_v(\omega).
\end{align}
Considering Eq.~\eqref{Eq:FDTNEq}, we can write the following alternative expresison
\begin{align}
  \label{Eq:ForceDipoleLate2.5}
  F
   &=
  \frac{\hbar}{\pi}
  \int_{0}^{\infty}\mathrm{d}\omega
  \int\frac{\mathrm{d}q}{2\pi}
  q
  \int \frac{\mathrm{d}\tilde{q}}{2\pi}
  \\\nonumber
  &\times
  \left\{
  \coth\left(\frac{\beta\hbar\omega_{q}^{+}}{2}\right)
  -
  \coth\left(\frac{\beta\hbar\omega_{\tilde{q}}^{+}}{2}\right)
  \right\}
  \\\nonumber
  &\times
  \mathrm{Tr}
  \left[
  \underline{\alpha}_v(\omega)
  \underline{G}_\Im(\tilde{q},\mathbf{R}_a,\omega_{\tilde{q}}^{+})
  \underline{\alpha}_v^{\dagger}(\omega)
  \underline{G}_{\Im}(q,\mathbf{R}_a,\omega_q^+)
  \right]
  ,
\end{align}
where we used that $2n(x)=\coth(x/2)-1$. Equation \eqref{Eq:ForceDipoleLate2.5} is the
result we would have obtained avoiding the frequency splitting and using the symmetric ordering for the operators (see Appendix \ref{App:DerForceGen}).
Similarly to what has already been observed in the zero-temperature case \cite{reiche20d}, the structure of the previous equation, quadratically in the Green tensor, reveals the strong non-additive features of the frictional interaction:
Modifying the geometry or material of the setup leads to a non-trivial modification of the viscosity coefficients \cite{reiche20d,reiche21,durnin21}.

It is instructive to consider first the simplifying assumption of Local Thermal Equilibrium (LTE).
The LTE approximation forgoes the opportunity to solve the system exactly in the steady-state [see Eq.~\eqref{Eq:FDTNEq}] and thereby ignores certain low-frequency contributions to the power spectrum of the interaction \cite{reiche20a}.
Instead, it directly applies the equilibrium fluctuation-dissipation theorem to the correlations of the dipole operator which technically amounts to replacing $n(\omega_q^+)\to n(\omega)$ in Eq.~\eqref{Eq:FieldSpectrum} (see Refs. \cite{intravaia2016a,reiche20a} for details) or equivalently $\coth(\beta\hbar\omega_{\tilde{q}}^{+}/2)\to \coth(\beta\hbar\omega/2)$ in Eq.~\eqref{Eq:ForceDipoleLate2.5}.
If we disregard the tensorial structure of the integral kernel (thereby neglecting rotational degrees of freedom \cite{intravaia2019}) and restrict our result to the leading order in $\alpha_0$, we restore the result of Refs. \cite{dedkov2017,oelschlaeger20}.
As explained above, a particular case of interest is the motion of the particle in the thermal field. Using the LTE approximation, we focus on the leading order in the velocity for which we have
$
\coth(\beta\hbar\omega_{q}^{+}/2)
-
\coth(\beta\hbar\omega/2)
\sim
-q\hbar\beta v/(2\sinh^2(\beta\hbar\omega/2))
$ \cite{estrada1994,estrada2002}. In the case of an empty space, we further find that
$\underline{\alpha}_{v,\Im}(\omega)=\mathrm{Im}[\alpha_v(\omega)]\mathbb{1}$ and
\begin{equation}
\int\frac{\mathrm{d}q}{2\pi}
q^2\mathrm{Tr}[\underline{G}_{\Im}(q,\mathrm{R}_a,\omega)]
=
\frac{\omega^5}{6\pi\epsilon_0c^5},
\end{equation}
with $\epsilon_0$ the vacuum permittivity \cite{tomas95}. In this case, Eq.~\eqref{Eq:ForceDipoleLate2.5} reduces to
\begin{align}
  \label{Eq:ForceDipoleLateVacuum}
  F^{\mathrm{LTE}}
  \sim
  -
  \frac{v\hbar^2 \beta}{3\pi c^5(4\pi\epsilon_0)}
  \int_{0}^{\infty}\mathrm{d}\omega~
  \mathrm{Im}[\alpha_{v=0}(\omega)]
  \frac{\omega^5}{\sinh^2\left(\frac{\beta\hbar\omega}{2}\right)},
\end{align}
which is similar to the result for black-body friction reported in Ref. \cite{mkrtchian2003} (see also Refs. \cite{maianeto04,lach12,lach12a,jentschura15,volokitin15,Kaur20}). Importantly, in Eq.~\eqref{Eq:ForceDipoleLateVacuum} the particle's optical response is considered to all orders \cite{jentschura15,volokitin15}.
As it should be expected for this case, the frictional force vanishes in the limit $T\to 0$ and the corresponding viscosity $\mu_T^{\mathrm{vac}}$ is exclusively thermal.
To make a first quantitative estimate, we can approximate the polarizability at the leading order atom-surface coupling, $\alpha_{0}$, as
$
\mathrm{Im}[\alpha_{v=0}(\omega)]\sim \mathrm{Im}[\alpha_{\rm B}(\omega)]= \alpha_0\omega_a\pi\delta(\omega_a-\omega)/2
$ ($\omega>0$) \cite{mkrtchian2003}.
At the leading order in the velocity, the vacuum thermal viscosity is then given by \cite{mkrtchian2003,lach12,volokitin15}
\begin{align}
\label{Eq:MuTVacSingleRes}
&\mu_T^{\mathrm{vac}}
  \sim
  -\frac{\alpha_0}{4\pi\epsilon_0}
  \frac{\hbar\omega_a^5}{3c^5}
  \frac{\beta \hbar\omega_a}{\sinh^2\left(\frac{\beta\hbar\omega_a}{2}\right)}.
\end{align}
The previous approximation is valid as long as $\mathrm{Im}[\alpha_{v=0}(\omega)]$ is sufficiently sharp (width much smaller than resonance frequency) with respect to the remaining part of the integrand of Eq.~\eqref{Eq:ForceDipoleLateVacuum} around $\omega \sim \omega_{a}$. This implies that $T\sim T_{a}\equiv\hbar\omega_a/k_{\mathrm{B}}$ (see also Sec. \ref{sec:noneqflu}).
In this case, one can speak of a resonant interaction were the thermal radiation has sufficient energy to excite the atomic transition and drive the absorption and the emission process at the leading order (single-excitation-) level. For optical transitions, however, this behavior occurs at the rather high temperatures of $\sim 10^{4}$ K.
At smaller temperatures, $k_{\mathrm{B}}T\ll\hbar\omega_a$, Eq.~\eqref{Eq:MuTVacSingleRes} does not represent the correct expression since low frequencies dominate the integrand in Eq.~\eqref{Eq:ForceDipoleVacHigh}.
At these energy, the interaction is non-resonant \cite{lach12}, and one needs to consider higher-order (two- or more-excitation-) processes to correctly describe the frictional dynamics \cite{intravaia2015b}.
The approximation we used to describe the polarizability for computing Eq.~\eqref{Eq:MuTVacSingleRes} is no longer adequate and we need to employ the dressed polarizability \cite{lach12}.
Upon inserting the vacuum Green tensor \cite{tomas95} in Eq.~\eqref{Eq:PolDressed}, we obtain at the leading order of the atom-surface coupling 
\begin{align}
\label{Eq:MuTVacSingleResLowT}
&\mu_T^{\mathrm{vac}}
\sim
  -\frac{32\pi^5}{135}\hbar
  \frac{\alpha_0^2}{\epsilon_0^2}\left(\frac{k_{\mathrm{B}}T}{\hbar c}\right)^8,
& k_{\mathrm{B}}T\ll\hbar\omega_a.
\end{align}
which replaces the exponentially damped result from Eq.~\eqref{Eq:MuTVacSingleRes}. The expressions in Eqs.~\eqref{Eq:MuTVacSingleRes} and \eqref{Eq:MuTVacSingleResLowT} have already been discussed in earlier works within different approaches \cite{lach12,jentschura15,volokitin15,guo21}.
%{\color{red} (different result for a nanoparticle also in comparison with \cite{mkrtchian2003})}.
A comparison of the last two expressions indicates the transition from a resonant regime dominated by $\omega\sim \omega_{a}$ to a non-resonant one where the thermal frequency $\omega_{\rm th}=k_{\mathrm{B}}T/\hbar$ becomes relevant.
Equation \eqref{Eq:MuTVacSingleResLowT} is one example of incorporating higher-order corrections of the atom-field coupling into the description and was discussed earlier in the context of one-loop corrections to the quantum-electrodynamical atom-field coupling \cite{jentschura15,volokitin15}.
The respective regimes of validity of the previous two equations can be connected to the nature of the available (dissipative) interaction channels. We will cover this topic with more detail in Section \ref{sec:noneqflu}.

Lastly, if we exchange the variables for the wave vector integration $q\leftrightarrow \tilde{q}$ for the second term in the integral kernel in Eq.~\eqref{Eq:ForceDipoleLate2} and consider the polarizability at the leading order in $\alpha_{0}$, i.e. $\underline{\alpha}_v(\omega)\sim\alpha_{\rm B}(\omega)$, we can write
\begin{align}
  \label{Eq:ForceDipoleLate3}
F&\sim
%\\\nonumber
%&\quad
-2
\frac{\hbar}{\pi}
\int_{0}^{\infty}\mathrm{d}\omega
\int\frac{\mathrm{d}q}{2\pi}
\int \frac{\mathrm{d}\tilde{q}}{2\pi}
n(\omega_{q}^{+})
\left\{
\tilde{q}
-
q
\right\}
\\\nonumber
&\quad\times \vert\alpha_{\rm B}(\omega)\vert^{2}
\mathrm{Tr}
\left[
\underline{G}_\Im(\tilde{q},\mathbf{R}_a,\omega_{\tilde{q}}^{+})
\underline{G}_{\Im}(q,\mathbf{R}_a,\omega_q^+)
\right]
.
\end{align}
This expression can be connected with the result found in Ref. \cite{guo21}. One important point is, however, that our result prescribes a distribution of the poles of the polarizability in the complex-frequency plane that is in agreement with causality and the optical theorem \cite{intravaia2011}. This feature is relevant for a correct evaluation of the resonant part of the interaction.

\section{Quantum and Thermal Viscosity}
\label{sec:quacacross}
We now turn our attention back to our main result in Eq.~\eqref{Eq:ForceDipoleLate2} and focus on the behavior of the viscosity coefficient  $\mu$ [see Eq.~\eqref{Eq:ViscosDef}].

Expanding the $\coth$-functions to the linear order in $v$, we obtain an electromagnetic viscosity coefficient which only depends on the temperature
\begin{align}
  \label{Eq:ForceDipoleVacHigh}
  \mu_{T}
 &\sim
  -\frac{\hbar^2\beta}{2\pi}
  \int_{0}^{\infty}\mathrm{d}\omega
  \int\frac{\mathrm{d}q}{2\pi}
  \int \frac{\mathrm{d} \tilde{q}}{2\pi}~
  \frac{q(q-\tilde{q})}
  {\sinh^2\left(\frac{\beta\hbar\omega}{2}\right)}
  \\\nonumber
  \times &
  \mathrm{Tr}
  \left[
  \underline{\alpha}_{v=0}(\omega)
  \underline{G}_\Im(\tilde{q},\mathbf{R}_a,\omega)
  \underline{\alpha}_{v=0}^{\dagger}(\omega)
  \underline{G}_{\Im}(q,\mathbf{R}_a,\omega)
  \right].
\end{align}

We start with a remark about the case of an atom moving through thermal vacuum.
Since the vacuum Green tensor is symmetric and even in $q$ and $\tilde{q}$ \cite{tomas95}, the term $\propto q\;\tilde{q}$ in Eq.~\eqref{Eq:ForceDipoleVacHigh} vanishes under the integration over the wave vectors.
Taking Eq.~\eqref{Eq:FDTSecondKind} into account, we recover Eq.~\eqref{Eq:ForceDipoleLateVacuum} and the corresponding vacuum polarizability, which was obtained employing the LTE approximation.
In other words, at the leading order in the velocity, the expression for the motion-induced electromagnetic viscosity acting on an atom in a vacuum at finite temperature coincides with the LTE approximation up to all orders in the atom-field coupling.

Equation \eqref{Eq:ForceDipoleVacHigh} becomes more interesting when the atom's environment is no longer homogeneous, i.e. when, in addition to the thermal field, other macroscopic objects are present in the space near the atom. In this case, the Green tensor is not necessarily symmetric \cite{tai94,dennis03} and the previous considerations made for vacuum do not apply.
For simplicity, in the following, we additionally assume that all the translationally invariant objects comprising the electromagnetic environment are made from an (not necessarily the same) Ohmic material \cite{jackson1999}.
Generalizations are possible but they require a more involved limiting procedure. As an example, we refer to Ref. \cite{oelschlager2018}, discussing the frictional interaction at zero temperature between an atom and lossy multilayer structures.
For our analyses, the Ohmic assumption mathematically translates into $\underline{G}_{\Im}'(q,\mathbf{R}_a,0)\neq0$, where the prime denotes a derivative with respect to frequency.

Inspecting Eq.~\eqref{Eq:ForceDipoleVacHigh} [or also Eq.~\eqref{Eq:ForceDipoleLate2}] one realizes that the thermal field relevantly impacts the interaction at frequencies $\omega\lesssim \omega_{\rm th}$.
At room temperature, this yields $\omega_{\rm th}\sim 26$ meV (about 0.34 meV at 4 K) which is much smaller than the typical transition frequencies of alkali metal atoms \cite{steck} or the resonances of common materials like metals (typically a few eV) \cite{barchiesi2014}.
For comparison, some dielectrics, such as silicon nitride or glass, feature a frequency behavior which can become comparable to thermal energies for temperatures $T$ reachable in experiments (e.g. the interface SiC/vcuum features a surface resonance around 120 meV \cite{joulain2005}). This was, for example, used in the context of radiative heat transfer \cite{mulet02,rousseau09}.
For our purposes it is hence reasonable to assume that the temperature cannot appreciably excite any resonance in the system.
As for Eq.~\eqref{Eq:MuTVacSingleResLowT}, our interest will then lie in the non-resonant part of the interaction, which provides the leading-order correction to the zero temperature result.

The leading terms in $T$ of Eq.~\eqref{Eq:ForceDipoleVacHigh} can be found by considering the moment asymptotic expansion \cite{estrada1994,estrada2002}
\begin{align}
\label{Eq:EstradaExpansionSinh}
\frac{\hbar\beta}{2\sinh^{2}\left(\frac{\beta\hbar\omega}{2}\right)}
    &\sim
  -
 2\delta(\omega)
  +
  \frac{\pi^2}{3}
  \frac{2}{\hbar^2\beta^2}
  \delta''(\omega),
\end{align}
where the double-prime denotes the second derivative with respect to frequency. The first term corresponds to the zero-temperature limit while the second term is the first thermal correction.
Upon inserting the above expression into Eq.~\eqref{Eq:ForceDipoleVacHigh}, we find, however, that the expression arising from the first term on the r.h.s. of Eq.~\eqref{Eq:EstradaExpansionSinh} always
identically vanishes since
$\underline{G}_{\Im}(q,\mathbf{R}_a,0)=0$ due to the crossing relation \cite{jackson1999}.
In other words, in the steady state, we always have $\lim_{T\to 0}\mu_{T}=0$, which is equivalent to say that at zero temperature the frictional force does not have a linear dependence in velocity \cite{dedkov2002,volokitin2007,viotti19}.
In empty space, this can be understood as a consequence of the system's Lorentz invariance.
When material interfaces are present, this result is less evident and has been discussed elsewhere \cite{scheel2009,intravaia2014,milton2016,klatt2021}.
In agreement with some earlier work \cite{intravaia2014,intravaia2015b,intravaia2016a,viotti19,reiche20d}, it indicates that in the steady state at $T=0$, when only quantum fluctuations are present, we would have to expand the frictional force in Eq.~\eqref{Eq:ForceDipoleLate2.5} to higher orders in $v$ to have a nonzero electromagnetic viscosity. In particular, in the steady state for the Ohmic case, quantum friction corresponds at the leading order in the velocity to the electromagnetic viscosity $\mu(v,0)=\mu_{\mathrm{QF}}\propto v^2$.
The latter has been analyzed in earlier works and
for completeness we report the result for $\mu_{\mathrm{QF}}$ at leading order in the atom-surface coupling,
\begin{align}
\label{Eq:muQF}
\mu_{\mathrm{QF}}
  \sim&
  -
  \frac{\hbar}{\pi}\alpha_0^2 v^{2}
  \int\frac{\mathrm{d}q}{2\pi}
  \int\frac{\mathrm{d}\tilde{q}}{2\pi}
  \frac{(q+\tilde{q})^4}{12}
  \\\nonumber
  &\times
  \mathrm{Tr}
  \left[
  \underline{G}'_{\Im}(\tilde{q},\mathbf{R}_a,0)
  \underline{G}'_{\Im}(q,\mathbf{R}_a,0)
  \right].
\end{align}
We recall that the previous expression is the result of an analysis showing that quantum friction at low velocities is typically a low-frequency phenomenon, dominated by $\omega\lesssim v/\lambda_{a}$, where $\lambda_{a}$ usually correspond to the minimal distance from one of the objects shaping the electromagnetic environment of the atom.
We remark that, once again, for Eq.~\eqref{Eq:muQF} it was assumed that $v/\lambda_{a}$ is much smaller than any of the system's resonance frequencies (note that for $v\sim 1$ km/s and $\lambda_{a}\sim 1$ nm, one has $v/\lambda_{a}\sim 0.6$ meV \cite{intravaia2016,reiche20d}).

The second term on the r.h.s. of Eq.~\eqref{Eq:EstradaExpansionSinh} gives the leading order temperature correction of the low-velocity electromagnetic viscosity in Eq.~\eqref{Eq:ForceDipoleVacHigh}.
In line with other analyses \cite{dedkov2002,volokitin2002,volokitin2007,dedkov2017,viotti19}, it scales quadratically in $T$ and corresponds to a frictional force that grows linearly with $v$ in the steady-state.
Focusing on the leading order in atom-surface coupling, using the symmetries of the integral kernel with respect to the wavevector as well as that $\underline{G}_{\Im}$ is odd with respect to frequency, the temperature correction can be written in a form which is similar to Eq.~\eqref{Eq:muQF}, i.e.
\begin{align}
  \label{Eq:ForceDipoleMat2}
  \mu_T
  \sim
  &-
  \alpha_0^2
  \frac{\pi}{3\hbar\beta^2}
  \int\frac{\mathrm{d}q}{2\pi}
  \int \frac{\mathrm{d}\tilde{q}}{2\pi}~
  (q-\tilde{q})^2
  \\\nonumber
  &\times
  \mathrm{Tr}
  \left[
  \underline{G}_{\Im}'(\tilde{q},\mathbf{R}_a,0)
  \underline{G}_{\Im}'(q,\mathbf{R}_a,0)
  \right].
\end{align}
It is important to stress that Eq.~\eqref{Eq:ForceDipoleMat2} takes into account both the symmetric and the skew-symmetric parts of the Green tensor as well as the backaction from the electromagnetic field onto the atom's dynamics \cite{volokitin2002,jentschura15,jentschura16}.
In particular, for a generic system that is translationally invariant along the $x$-axis, we can write the Hermitian tensor $\underline{G}_{\Im}(q,\mathbf{R}_a,\omega)$ in terms of a real, symmetric matrix $\underline{\Sigma}(q,\mathbf{R}_a,\omega)$ (even in $q$) and a real vector $\mathbf{s}_{\perp}(q,\mathbf{R}_a,\omega)$ (odd in $q$) normal to the invariance axis. We have indeed that \cite{reiche20d}
\begin{equation}
\label{GSplit}
\underline{G}_{\Im}(q,\mathbf{R}_a,\omega)=\underline{\Sigma}(q,\mathbf{R}_a,\omega)+\mathbf{s}_{\perp}(q,\mathbf{R}_a,\omega)\cdot\mathbf{\underline{L}},
\end{equation}
where $[\underline{L}_{i}]_{jk}=-\ii \epsilon_{ijk}$ denotes the generator of 3D rotations around the $i$-axis [see also discussion below Eq.~\eqref{eq:flucdisE0}].
The vector $\mathbf{s}_{\perp}(q,\mathbf{R}_a,\omega)$ is connected to the spin-dependent part of the electromagnetic density of states (see Refs. \cite{dennis03,bliokh2015,intravaia2019} for details).
Using the properties of the trace, Eq.~\eqref{GSplit} leads to the decomposition
\begin{align}
\label{Eq:MuSplit}
\mu_T
  &=
  \mu_T^{t}+\mu_T^r.
\end{align}
The first term contains only the symmetric matrix $\underline{\Sigma}$
\begin{align}
\mu_T^t
  &
  \sim
  -\alpha_0^2
  \frac{2\pi}{3\hbar\beta^2}
  \int\frac{\mathrm{d}q}{2\pi}
  \int \frac{\mathrm{d}\tilde{q}}{2\pi}~
  q^2
  \\\nonumber
  &\qquad\times
  \mathrm{Tr}
  \left[
  \underline{\Sigma}'(\tilde{q},\mathbf{R}_a,0)
  \underline{\Sigma}^{\prime}(q,\mathbf{R}_a,0)
  \right]
\end{align}
and its expression can be related to the description of the thermal correction of the electromagnetic viscosity which is usually obtained within the LTE approximation.
Using $\mathrm{Tr}[\underline{L}_{i}\underline{L}_{j}]=2 \delta_{ij}$, the second term reads
\begin{align}
\mu_T^r
  &
  \sim
  \alpha_0^2
  \frac{4\pi}{3\hbar\beta^2}
   \int\frac{\mathrm{d}q}{2\pi}
  \int \frac{\mathrm{d}\tilde{q}}{2\pi}~
  q\;\tilde{q}
  \\\nonumber
  &\qquad\times
  \mathbf{s}_{\perp}'(\tilde{q},\mathbf{R}_a,0)\cdot
  \mathbf{s}_{\perp}'(q,\mathbf{R}_a,0)
  .
\end{align}
It is connected to the interaction between atom and electromagnetic excitation with nonzero spin and, to the best of our knowledge, its role has been highlighted only for quantum friction \cite{intravaia2019} but not for the finite temperature case.
Interestingly, like for the zero-temperature limit, an analysis shows that $\mu_T^r$ tends to decrease the viscosity experienced by the moving atom (e.g. see Eq.~\eqref{Eq:muTtmuTr} below). In combination with the spin-momentum locking this reduction can be related to an exchange and a net transfer of angular momentum from the field to the atom \cite{intravaia2019}.
Given its physical origin, the impact of $\mu_T^r$ can be diminished choosing an axis-symmetric geometry that, for a trajectory along the symmetry axis, suppresses the exchange of angular momentum between the atom and the radiation  \cite{reiche20d}.

\begin{figure}[t]
  \includegraphics[width=\linewidth]{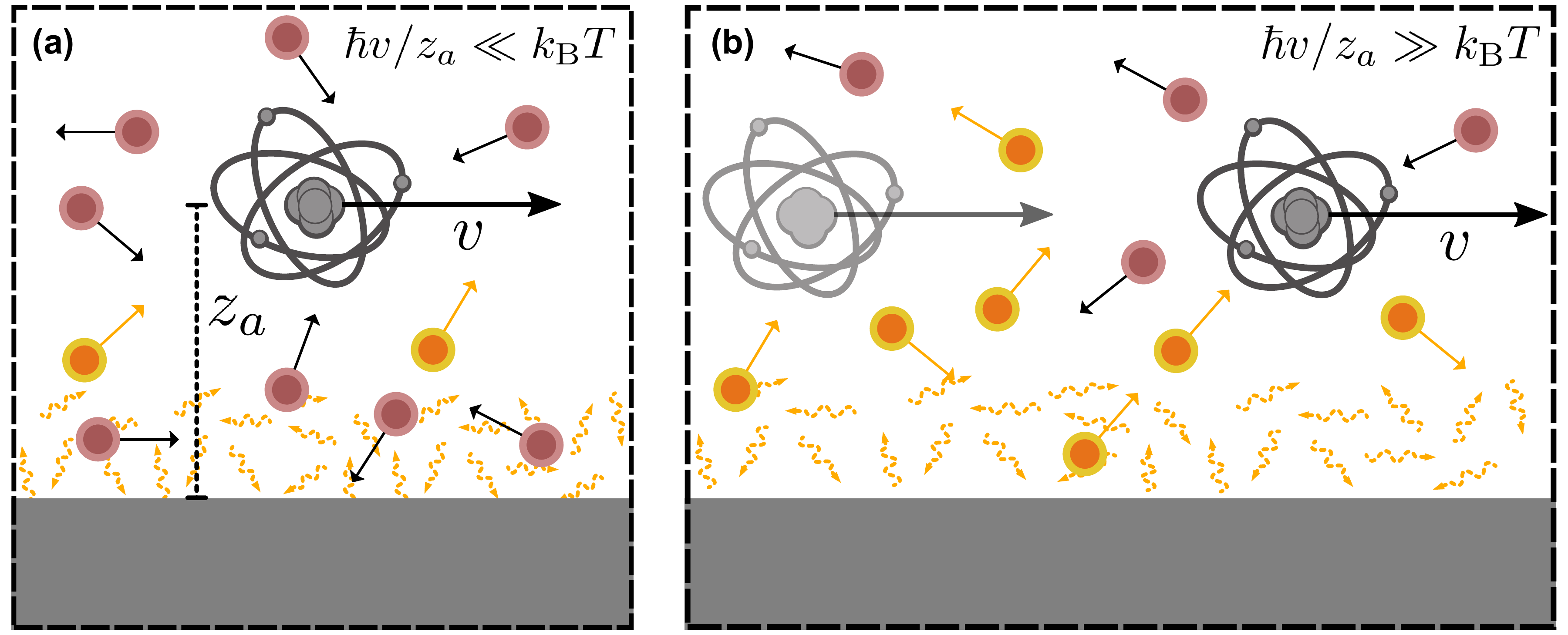}
  \caption{\label{fig:mechanisms}
  Schematic of the different mechanisms at work for the determination of the viscosity on an atom
  moving in a complex structured electromagnetic environment (see main text for more details).
  \textbf{(a)} At sufficiently low velocities
  large distances or high temperatures the frictional process is dominated by thermal excitations (red points with black arrows).
  \textbf{(b)} At high velocities, short separations or low temperatures the system behaves more
  according to its quantum characteristics: The viscous dynamics is
  determined by the interactions with the electromagnetic quantum fluctuations. Due to the anomalous-Doppler-effect
  virtual excitations can become real at the expense of the kinetic energy of the atom (yellow points with yellow arrows). The relevance of these
  excitation ``extracted'' from vacuum grows with $v$ and dominates the thermal interaction at sufficiently high
  velocities.
  In both limits, the local electromagnetic density of states is modified by the presence of the vacuum-material interface (yellow curly arrows), inducing a functional dependence of the frictional interaction on the atom-surface separation ($\lambda_{a}=z_{a}$).
 }
\end{figure}

Lastly, we would like to point out that, while $\mu_T\to\mu^{\rm vac}_T\propto T^8$ in Eq.~\eqref{Eq:MuTVacSingleResLowT} is related to the general properties of vacuum, the quadratic temperature dependence of $\mu_T$ in Eq.~\eqref{Eq:ForceDipoleMat2} is connected to the presence of objects and our Ohmic assumption on the electromagnetic response of the involved materials. This is also equivalent to a local density of states that scales linearly with the frequency for $\omega\to 0$. Deviations from the Ohmic behavior, including those leading to $\mu^{\rm vac}_T$, are included in our theory, but require to adequately consider Eq. \eqref{Eq:ForceDipoleVacHigh} and its low frequency limit.
Mathematically, these features are encoded in the Green tensor $\underline{G}$. In particular, the expected transition $\mu_T\propto T^2 \to T^{8}$ as a function of the distance of the atom from the objects can be seen as a consequence of the general decomposition $\underline{G}=\underline{G}_{0}+\underline{G}_{\rm sc}$. Here, $\underline{G}_{0}$ is the homogeneous vacuum Green tensor and $\underline{G}_{\rm sc}$ is the scattering part of Green tensor, due to the presence of the objects \cite{intravaia11a}. The two tensors not only scale differently with the frequency but also with
$\mathbf{R}_{a}$, describing the position of the atom: while $\underline{G}_{0}$ does not depend on $\mathbf{R}_{a}$ and is always nonzero, the contribution $\underline{G}_{\rm sc}$ scales as the inverse of the distance of the atom to the objects around it ($\sim \lambda_{a}$), making the scattering contribution more or less relevant as a function of $\mathbf{R}_{a}$. Some rough but general considerations can also be made regarding the transition $\mu_{\rm QF}\to \mu_T\propto T^2$. As discussed above, both viscosity coefficients are connected with the low-frequency behavior of the integrand in Eq.~\eqref{Eq:ForceDipoleLate2.5}. The two characteristic frequency-scales are, respectively,
$v/\lambda_{a}$ and $k_{\rm B}T/\hbar$. In the Ohmic limit, the larger is the frequency range, the stronger is the effect, indicating that
$\mu_{\rm QF}\geq \mu_T$ if $v/\lambda_{a}\geq k_{\rm B}T/\hbar$ and vice versa.
This allows us to define the following critical velocity, distance and temperature
\begin{equation}
\label{criticalScales}
v_{c}\sim \frac{k_{\rm B}T\lambda_{a}}{\hbar},
\quad \lambda_{c}\sim \frac{\hbar v}{k_{\rm B}T},
\quad T_{c}\sim \frac{\hbar v}{k_{\rm B}\lambda_{a}},
\end{equation}
where $\mu_{\rm QF}$ and $\mu_T$ interchange their role as dominant contribution to the interaction, emphasizing which kind of phenomenon (thermal vs quantum fluctuations) is responsible for the breaking of Lorentz invariance. Curiously, the atom needs to have a sufficiently high kinetic energy ($v>v_{c}$) to start to capture more of the ``quantumness'' of the frictional force  (at $T=3$ K and for $\lambda_{a}\sim 10$ nm one has that $v_{c}\sim 4$ km/s). Indeed, for $v<v_{c}$ the electromagnetic viscosity is essentially given by $\mu_T$.
Physically, this behavior can be understood by considering the different mechanisms in the frictional process.
As for black-body friction, it is the thermal bath that non-resonantly drives the interaction at low velocity [see Fig.~\ref{fig:mechanisms} (a)].
However, due to the boundary conditions imposed by the non-homogeneity of the environment, the density of states for the electromagnetic field is substantially different from that of the empty space, inducing a modification in the interaction which depends on the position of the atom (see also Sec. \ref{sec:noneqflu} below).
Simultaneously, however, due to the motion-induced anomalous-Doppler-effect \cite{ginzburg1960,nezlin1976,maghrebi2013,intravaia2019}, \emph{virtual} excitations can be ``energized'' at the expense of the atomic kinetic energy and become \emph{real}, participating in the frictional process [see Fig.~\ref{fig:mechanisms} (b)]. Their relevance is proportional to the velocity of the atom and connected to the shape of density of states of the electromagnetic field, determining the functional dependency in $v$ of $\mu_{\rm QF}$. Therefore, for constant $T$, depending on $v$, either the quantum or the thermal mechanism dominate the frictional interaction.

In order to illustrate our general findings in more detail, it is instructive to consider a concrete geometry and compute the viscosity coefficients for an atom moving in vacuum with respect to a single planar half space located at $z\leq 0$ and comprised from a local, Ohmic, dispersive and dissipative material. In this case $z_{a}=\lambda_{a}$ measures the distance from the bulk's interface.
For simplicity, we focus here on a motion in the near-field of the surface since this is the regime where the frictional interaction becomes most relevant \cite{dedkov2017}.
Upon inserting the expression for the Green tensor \cite{tomas95,intravaia2016a,intravaia2019} and performing the wave vector integrals, we find \cite{volokitin2002,volokitin06,jentschura16}
\begin{align}
\label{Eq:muTtmuTr}
&\mu_T
  =
  -\frac{3}{\pi}\hbar\alpha_0^2\rho^2\frac{\left(\frac{k_{\mathrm{B}}T}{\hbar}\right)^2}{(2z_a)^{8}},
&\mu_T^r=-\frac{\mu_T^t}{2},
\end{align}
where in general $\rho=\lim_{\omega\to 0}\partial_{\omega} \mathrm{Im}[r(\omega)]/(2\epsilon_{0})$ with $r(\omega)$ the transverse magnetic Fresnel coefficient.
The parameter $\rho$, which is also connected to the low-frequency tail of the local density of states, highlights the dissipative properties of the electromagnetic environment and, specifically for conductors, effectively corresponds to the resistivity of the involved material.
Notice that, since it is connected with our assumption of an Ohmic behavior, the limit $\rho\to 0$ in the previous expression has to be handled with care and in general does not imply the vanishing of the electromagnetic viscosity.
We already saw that, when $\rho$ vanishes because no objects are present ($\underline{G}_{\rm sc}\to 0$), Eq.~\eqref{Eq:muTtmuTr} needs to be replaced by Eqs.~\eqref{Eq:MuTVacSingleRes} and \eqref{Eq:MuTVacSingleResLowT}.
For non-Ohmic materials, Eq.~\eqref{Eq:muTtmuTr} needs to be reevaluated.
For example, in the limit of a non-dissipative material with a constant and real refraction index
$n$, the frictional interaction is occurring above the Cherenkov threshold, i.e. $v\gtrsim c/n$ \cite{pieplow2015,maghrebi2013}.
In the specific geometry considered here, the spin-sensitive parts of the interaction turn out to reduce the viscosity by a factor of one half.
For comparison, for the same setup, the electromagnetic viscosity exclusively due to quantum fluctuations is also given by $\mu_{\rm QF}=\mu^{t}_{\rm QF}+\mu^{r}_{\rm QF}$, where
\begin{align}
\label{Eq:muQFtmuQFr}
&\mu_{\rm QF}
  =
  -\frac{18}{\pi^{3}}\hbar\alpha_0^2\rho^2\frac{v^2}{(2z_a)^{10}},
&\mu^r_{\rm QF}=-\frac{5}{7}\mu_{\rm QF}^t.
\end{align}
This corresponds to a reduction of of $\mu_{\rm QF}^t$ by roughly 70\% due to $\mu_{\rm QF}^r$ \cite{intravaia2019}.

%\begin{comment}

\begin{figure}
   \includegraphics[width=\linewidth]{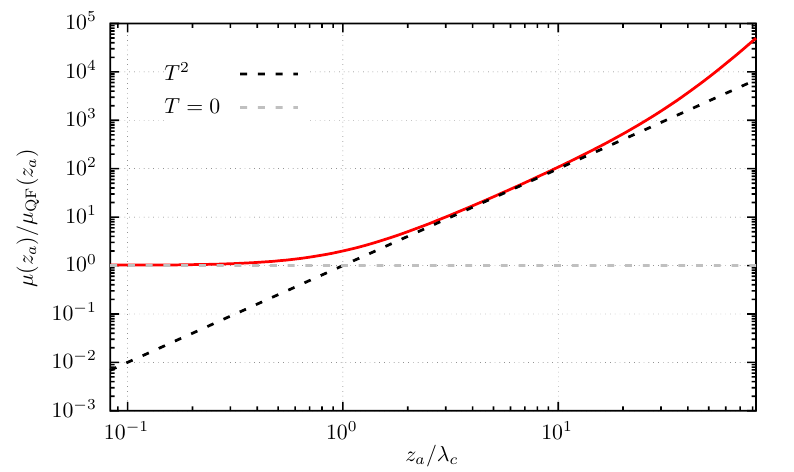}
	\caption{Electromagnetic viscosity of an 
%	$^{87}$Rb 
	atom moving near a planar interface
($\alpha_{0}= 4\pi \epsilon_{0} \times 47.28$ \AA$^{3}$, $\omega_{a}=\fran{1.3}$ eV 
%\cite{steck}
).
The surface is described by a Drude-permittivity  
$\epsilon(\omega)=1-\omega_{\rm p}^{2}/[\omega(\omega+\imath \Gamma)]^{-1}$
using the parameters $\omega_{\rm p}=9$ eV, $\Gamma=100$ meV, giving
$\rho=\Gamma/[\epsilon_{0}\omega_{\rm p}^{2}]=9.18\times 10^{-8}$ $\Omega$m.
The plot depicts $\mu=F/v$ [(full red line) see e.g. Eq.~\eqref{Eq:ForceDipoleLate}] as a function of the distance $z_{a}$ 
at a constant temperature $T=3$ K and velocity $v=12$ km/s. The viscosity is normalized with respect to the 
asymptotic expression $\mu_{\rm QF}$ given in Eq~\eqref{Eq:muQFtmuQFr}. At
separations $z_{a}/\lambda_{c}<1$ [see Eqn.~\eqref{ratio}] the drag is essentially due to quantum friction (horizontal dashed grey line). At distances $z_{a}/\lambda_{c}>1$ the electromagnetic viscosity $\mu_{T}\propto T^{2}$ becomes dominant (black dash line).
Black-body friction starts to be relevant at separations $z_{a}/\ell_{c}>1$ [see Eqs.~\eqref{ratio}].
\label{fig:za}}
\end{figure}

\begin{figure}
 \includegraphics[width=\linewidth]{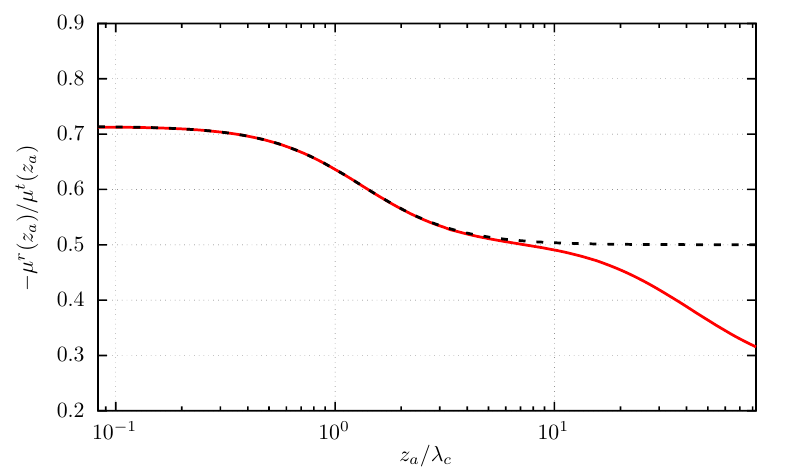}
	\caption{Behavior of the electromagnetic viscosities components $\mu^{r}$ and $\mu^{t}$ as a function of the 
	%(a)  
	atom-surfaces separation for $T=3$ K and $v=12$ km/s (full red line). The atom and the surface are described in terms of the same parameters given in Fig.~\ref{fig:za}.
	%and (b) of the system's temperature for $z_{a}=$. 
	%Both components are normalized to the asymptotic expression for $\mu^{t}$ [see Eq.~\eqref{}]
	The dashed black line describes the asymptotic behavior predicted according to Eqs.~\eqref{Eq:muTtmuTr} and \eqref{Eq:muQFtmuQFr}, where the contribution of black-body friction was neglected.
The plot shows that the transition from $-\mu^{r}/\mu^{t}\sim 5/7$ to  $-\mu^{r}/\mu^{t}\sim 1/2$ occurs around the characteristic distance $z_{a}\sim\lambda_{c}$. For $z_{a}>\ell_{c}$, the viscosity enters the regime where $\mu_{T}^{\rm vac}$ becomes dominant. For these distances $\mu^{r}$ goes to zero.
%and temperature $T_{c}$.
%	\\
%	\dan{Caption to be adjusted depending on the plots.}
	\label{Fig:MuTMuR} }
	\end{figure}

Defining for simplicity the reduced thermal wavelength $\lambdabar_{\rm th}=c/\omega_{\rm th}=\hbar c/(k_{\rm B}T)$ ($\sim 7.6\;\mu$m at 300 K) and $\lambda_{\rho}=4\pi\epsilon_0 c\rho$ ($\sim 3\;$nm for $\rho=9.18\times 10^{-8}$ $\Omega$m), the expressions given above allow for a direct evaluation of the ratios
($k_{\mathrm{B}}T\ll\hbar\omega_a$)
\begin{subequations}
\label{ratio}
\begin{align}
& \frac{\mu_T}{\mu_{\mathrm{QF}}}
  \sim
  \frac{z_a^{2}}{\lambda_{c}^{2}}, 
  \quad \lambda_{c}\equiv\sqrt{\frac{3}{2\pi^{2}}}\frac{v}{c}\lambdabar_{\rm th}~,
  \\
&\frac{\mu^{\rm vac}_{T}}{\mu_{T}}
  \sim
  \frac{z_a^{8}}{\ell_{c}^{8}},
  \quad
  \ell_{c}\equiv\frac{\sqrt{2}}{4}\left[\sqrt{\frac{5}{2}}\frac{\lambda_{\rho}}{\lambdabar_{\rm th}}\right]^{\frac{1}{4}} \frac{c}{v}\lambda_{c}~.
\end{align}
\end{subequations}

\begin{figure}
  \includegraphics[width=\linewidth]{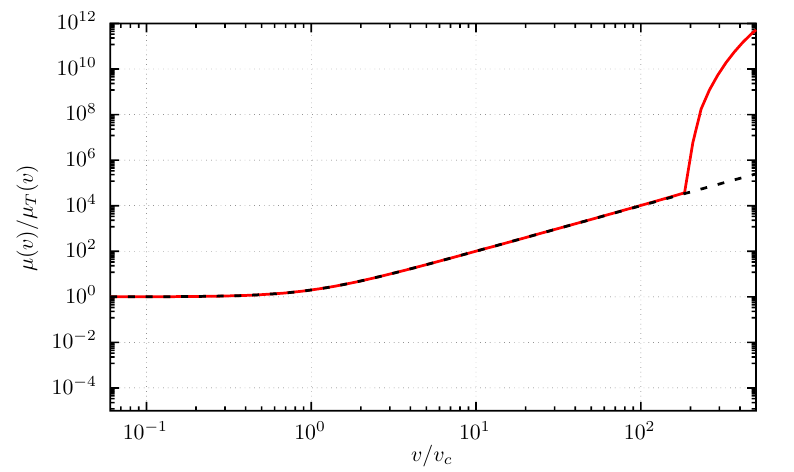}
	\caption{\label{fig:v1v3} Electromagnetic viscosity on an atom moving near a surface as a function of the velocity for constant temperature $T= 3$ K and distance $z_{a}=5$ nm. The atom and the surface are described in terms of the same parameters given in Fig.~\ref{fig:za}.
%	The distance is chosen in such a way that the vacuum contribution $\mu_T^{\mathrm{vac}}$ is subleading. 
	The function $\mu(v)$ (full red line) is normalized with respect to the asymptotic expression for $\mu_{T}$ given in Eq.~\eqref{Eq:muTtmuTr}. 
	At low velocity $v< v_{c}$ [see Eqn.~\eqref{additionalratio}] the viscosity is characterized by its thermal behavior, which for the chosen parameters is dominated by $\mu_{T}$. For $v>v_{c}$ the quantum component becomes relevant and the interaction behaves as $\mu_{\rm QF}\propto v^{2}$. 
	The dashed black line describes the non-resonant asymptotic behavior according to Eqs.~\eqref{Eq:muTtmuTr} and \eqref{Eq:muQFtmuQFr}.
	At sufficiently high velocities the viscosity enters the resonant regime, visible as a sudden rise of the viscosity as an function of $v$ (see Sec.~\ref{sec:noneqflu}). 
  }
\end{figure}

\begin{figure}
  \includegraphics[width=\linewidth]{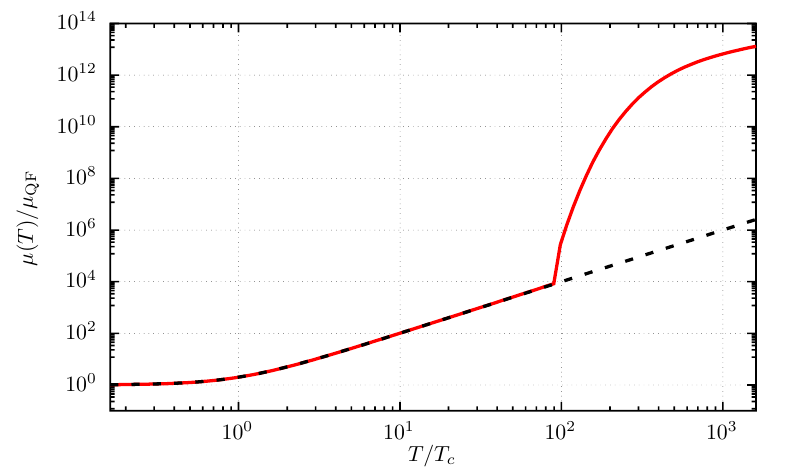}
	\caption{Plot of the viscosity $\mu$ as a function of the temperature for an atom moving near a planar surface. The atom and the material parameters are the same as in Fig.~\ref{fig:za}. The velocity and the distance from the surface are kept constant ($v=12$ km/s and $z_{a}=5$ nm). The value of $\mu(T)$ (full red line) is normalized with respect to $\mu_{\rm QF}$ given in Eq.~\eqref{Eq:muQFtmuQFr}. The dashed black line describes the asymptotic behavior according to Eqs.~\eqref{Eq:muTtmuTr} and \eqref{Eq:muQFtmuQFr}. At $T<T_{c}$ [see Eqn.~\eqref{additionalratio}] the frictional interaction is dominated by the quantum fluctuations of the system, while at higher temperatures it becomes more thermal and the viscosity is described by $\mu_{T}$. For temperature $T\gtrsim T_{a}$ the interaction becomes resonant and is characterized by an strong increases as a function of the temperature (see Sec.~\ref{sec:noneqflu}). \label{fig:T0T2}
  }
\end{figure}

We illustrate these findings in Figs. \ref{fig:za} -\ref{fig:T0T2}.
In Fig.~\ref{fig:za}, we depict the full expression for the electromagnetic viscosity $\mu$ [see e.g. Eq.~\eqref{Eq:ForceDipoleLate2.5}] at constant velocity and temperature as a function of the distance $z_{a}$ of the atom from the interface with the material.
As expected, the quantum frictional force dominates for  $z_{a}<\lambda_{c}$.
In the proximity of macroscopic surfaces comprised of common (dissipative) materials, both the viscosities $\mu_T$ and $\mu_T^{\mathrm{vac}}$ are usually subleading. At separations larger than $\lambda_{c}$, the viscosity connected to $\mu_{T}$ becomes relevant, while the effect of the distance-independent black-body friction occurs for $ z_{a}> \ell_{c}$, where $\mu^{\rm vac}_{T}/\mu_{T}>1$.
Figure \ref{Fig:MuTMuR} 
%(a) 
depicts the ratio $-\mu^{r}/\mu^{t}$ as a function of the atom-surface separation for the same parameters of Fig.~\ref{fig:za}. 
In general, as for the $T=0$ case \cite{intravaia2019}, $\mu^{t}$ can be defined starting from Eq.~\eqref{Eq:ForceDipoleLate} by retaining only the symmetric part of the Green tenor and of the tensor in the curly brackets.
Similarly, the expression for the viscosity $\mu^{r}$ can be derived from Eq.~\eqref{Eq:ForceDipoleLate} by considering the anti-symmetric parts of the same tensors. From Eq.~\eqref{GSplit} we have then that $\mu^{r}$ always identically vanishes if $\mathbf{s}_{\perp}=0$.
%In general, the $\mu^{t}$ can be defined from Eq.~\eqref{Eq:ForceDipoleLate} by using in all the quantities the tensor $\underline{\Sigma}(q,\mathbf{R}_a,\omega)$ instead of
%$\underline{G}(q,\mathbf{R}_a,\omega)$.
%The viscosity $\mu^{r}=\mu-\mu^{t}$ would then vanish if $\mathbf{s}_{\perp}=0$.
In agreement with our results in Eqs.~\eqref{Eq:muTtmuTr}  and \eqref{Eq:muQFtmuQFr}, we observe a partial cancellation of the two contributions to the viscosity ($-\mu^r/\mu^t$) that ranges from 70\% to 50\%, where the crossover takes place around $z_{a}/\lambda_{c}\sim 1$.
%Figure \ref{Fig:MuTMuR}  (b) shows the same ratio but for a constant distance from the surface and as a function of the temperature. The transition from $5/7$ to $1/2$ is visible for $T\sim T_{c}=\fran{\sqrt{3/(2\pi^{2})}} \hbar v/(k_{\rm B}z_{a})$.
%

For the atom-plate configuration, according to Eqs.~\eqref{ratio}, the behavior of the electromagnetic viscosity as a function of the velocity and of the temperature is characterized by the quantities
\begin{subequations}
\label{additionalratio}
\begin{gather}
v_{c}\equiv\sqrt{\frac{2}{3}}\pi \frac{z_a}{\lambdabar_{\rm th}},\quad T_{c}\equiv \sqrt{\frac{3}{2\pi^{2}}} \frac{\hbar v}{k_{\rm B}z_a }
, 
\\ \mathcal{T}_{c}\equiv\frac{\sqrt{2}}{8}\frac{c}{v}
\left(\frac{\sqrt{30}}{\pi}\frac{\lambda_{\rho}}{z_{a}}\right)^{\frac{1}{3}} T_{c}.
\end{gather}
\end{subequations}
As shown in Fig.~\ref{fig:v1v3}, the quantum frictional interaction starts to be relevant for the drag force acting on the particles only when $v> v_c$. At lower velocities the viscosity is instead characterized by $\mu_{T}$ (for sufficiently small distance we can neglect the impact of black-body friction). At sufficiently high velocities ($v\gtrsim \omega_{a} z_{a}$) the frictional interaction enters the resonant regime, where $\mu(v)$ sudden increases with the $v$ \cite{intravaia2016a} (see also Sec. \ref{sec:noneqflu}).
In Fig.~\ref{fig:T0T2} a similar behavior is visible as a function of the temperature. For $T<T_{c}$, the viscosity is dominated by the quantum fluctuations, while the thermal
effects appear at $T>T_{c}$ where $\mu(T)\sim \mu_{T}\propto T^{2}$. At a higher temperature $T\gtrsim T_{a}\equiv \hbar \omega_{a}/k_{\rm B}$, the interaction becomes resonant and grows faster with $T$ (see Sec.~\ref{sec:noneqflu}). Black-body friction becomes relevant for $T> \mathcal{T}_{c}$. 
%For the parameters chosen in Fig.~\ref{fig:T0T2} we have that $\mathcal{T}_{c}\sim 4.5 \times 10^{3}\;T_{c}$ and therefore the impact of black-body friction is not shown.

\begin{figure}[t]
  \includegraphics[width=\linewidth]{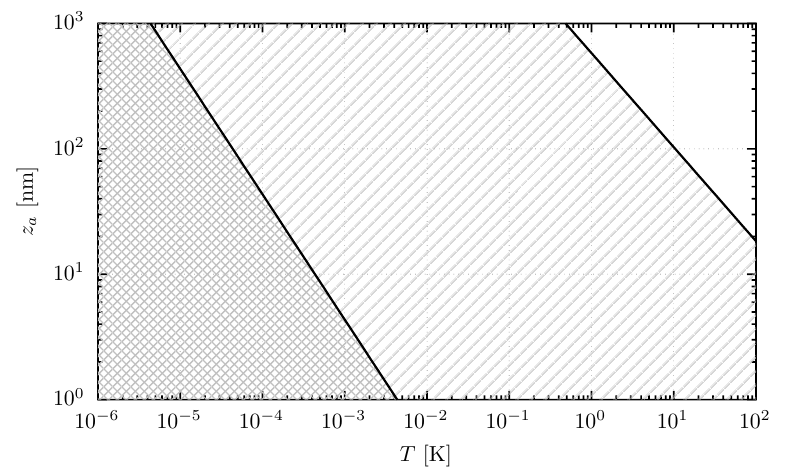}
	\caption{\label{fig:z_T}
  Physical character of the electromagnetic viscosity at constant velocity as a function of distance $z_a$ and temperature $T$.
    We can clearly distinguish three regimes. (i) For small separations or small temperatures (low-left corner of the plot) the quantum frictional viscosity $\mu\sim \mu_{\rm QF}$ and the system's quantum behavior dominate the frictional interaction.
  (ii) At $z_a T\geq \sqrt{3/2}\hbar v/(k_{\mathrm{B}}\pi)$, the thermal fluctuations from the material interface $\mu_T$ take over the main role of interaction.
  (iii) Finally, for distances and temperatures in the upper-right corner, at
  $
  z_a T
  \geq  [(9/2)\sqrt{10}\lambda_{\rho}/\lambda_{\rm th}]^{1/4} \hbar c/(4\pi k_{\rm B})
  $, black-body friction $\mu_T^{\mathrm{vac}}$ prevails over the surface contributions.
The full-blue line gives the $(z_a,T)$-values for which $\mu_{\mathrm{QF}}=\mu_T$, while the dashed-orange describes the case where $\mu_T=\mu_T^{\mathrm{vac}}$ [see Eqs. \eqref{ratio}].
  For our numerical example, we use the critical velocity $v=v_c\equiv\sqrt{2/3} \pi k_{\rm B} T z_a/\hbar$ and once the parameters given in Fig.~\ref{fig:za}.
%  again a $^{87}$Rb atom \cite{steck} moving in the close vicinity of a planar (local) gold surface \cite{tomas95,barchiesi2014}.
}
\end{figure}

Finally, figure \ref{fig:z_T} describes the relevance of the different contributions to the electromagnetic viscosity for given velocity in the $(z_{a}, T)$-plane.
While the quantum characteristic of the system ($\mu_{\rm QF}$) is predominant for parameters in the lower-left corner of this plane, the thermal effects start to be more relevant in the remaining part of the plane ($\mu_{T}$), and eventually recover the case of black-body friction ($\mu_{T}^{\rm vac}$) in the upper-right corner.

%\end{comment}

%
%
% NONEQUILIBRIUM FLUCTUATIONS
%
%
\section{Material-Modified Spectral Density}
\label{sec:noneqflu}
It is also interesting to inspect the behavior of $\mu_{T}$ from another perspective. In the simplest case, where the particle moves through thermal (homogeneous and isotropic) vacuum and no macroscopic bodies are present, we can rewrite Eq.~\eqref{Eq:ForceDipoleLateVacuum} in terms of the thermal part of Planck's black-body spectrum (the zero-point fluctuations are not relevant for the following considerations)
$
\varrho(\omega)
  =\hbar\omega^3 n(\omega)/(\pi^2 c^3)
$, i.e.
\begin{align}
\label{Eq:MuVacuum}
  \mu_T^{\mathrm{vac}}
  \sim
  -
  \int_{0}^{\infty}\mathrm{d}\omega~
  \frac{\omega\,
  \mathrm{Im}[\alpha_{v=0}(\omega)]}{3 c^2\epsilon_0}
  ~
  T\partial_T\varrho(\omega)
  ,
\end{align}
which can be shown to be related to Einstein and Hopf's original result \cite{einstein10a,lach12a}.
At given frequency $\omega$, the friction is connected to the slope of the Planck distribution $T\partial_T\varrho(\omega)$ at temperature $T$, which
is accounting for the Doppler-shift of the thermal field that the moving atom perceives \cite{lach12a}. The resulting distribution has a maximum and a variance both of the order of the thermal frequency $\omega_{\rm th}$.
Importantly, the atom only perceives the part of the spectrum that is within the frequency range of its interaction channels.
For instance, if we again consider a single dipole resonance [see Eq.~\eqref{Eq:MuTVacSingleRes} and discussion above], the usually sharp atomic transition acts as a filter, selecting only frequencies $\omega\sim\omega_a$. Mathematically, this is represented in Eq.~\eqref{Eq:MuVacuum} by the spectral density kernel $\eta^{\mathrm{vac}}(\omega)\equiv\omega\,
\mathrm{Im}[\alpha_{v=0}(\omega)]/(3 c^2\epsilon_0)$.
The interaction is therefore determined by the kind and the strength of the overlap between the spectral density filter and the temperature-dependent behavior of the electromagnetic field. The relation between the maxima and the widths of $\kappa^{\mathrm{vac}}(\omega)$ vs. $T\partial_T\varrho(\omega)$ determines which of the approximations in Eqs.~\eqref{Eq:MuTVacSingleRes} or \eqref{Eq:MuTVacSingleResLowT} is the most adequate.

We can extend this representation to more generic situations and write the thermal viscosity as
\begin{align}
  \label{Eq:ForceDipoleFilter}
  &
  \mu_T
\sim
  -
  \int_{0}^{\infty}\mathrm{d}\omega
  ~
  \eta(\omega)
  ~
  T\partial_T\varrho(\omega)
  .
\end{align}
Comparing to the expression in Eq.~\eqref{Eq:ForceDipoleVacHigh}, we can define the spectral density of the joint atom+field system as
\begin{align}
\label{Eq:RhoEM}
& \eta(\omega)
  =
  \frac{2\pi c^3}{\omega^4}
\int\frac{\mathrm{d}q}{2\pi}
\int \frac{\mathrm{d}\tilde{q}}{2\pi}~
q(q-\tilde{q})
\\\nonumber
&\quad\times
\mathrm{Tr}
\left[
\underline{\alpha}_{v=0}(\omega)
\underline{G}_\Im(\tilde{q},\mathbf{R}_a,\omega)
\underline{\alpha}_{v=0}^{\dagger}(\omega)
\underline{G}_{\Im}(q,\mathbf{R}_a,\omega)
\right],
\end{align}
which incorporates the interaction with the material-modified electromagnetic field up to all orders in coupling.
In this generalization, both the particle's polarizability and the Green tensor of the environment feature characteristic material-dependent resonances that manifest themselves as peaks in the behavior of $ \eta(\omega)$, singling out certain frequencies in the interaction with the thermal field.
If the temperature is sufficiently high, so that $\omega_{\rm th}$ is larger than the resonances' widths which $T\partial_T\rho(\omega)$ is enclosing, the interaction is resonant and an approximation similar to that in Eq.~\eqref{Eq:MuTVacSingleRes} is possible.
Since some of the resonances' widths scale as the inverse of the atom-surface separation, the resonant regime can also take on greater significance for larger distances from the surface.
The higher the temperature (or the larger the distance from the surface), though,  the more pronounced becomes also the impact of the vacuum viscosity $\mu^{\mathrm{vac}}_T$ which is independent from the surfaces [see e.g. Eq.~\eqref{Eq:MuTVacSingleRes}].
Conversely, at low temperature, $T\partial_T\rho(\omega)$ is significantly nonzero at low frequencies, acting  as a low-frequency bandpass filter where the dissipative features of the system dominate.
In this case the viscosity given by the integral in Eq.~\eqref{Eq:ForceDipoleFilter} is in the non-resonant regime which we analyzed in Sec. \ref{sec:quacacross}

%\begin{comment}
%%%%%%%%%%%%%%%%%%
\begin{figure}[t]
 \includegraphics[width=\linewidth]{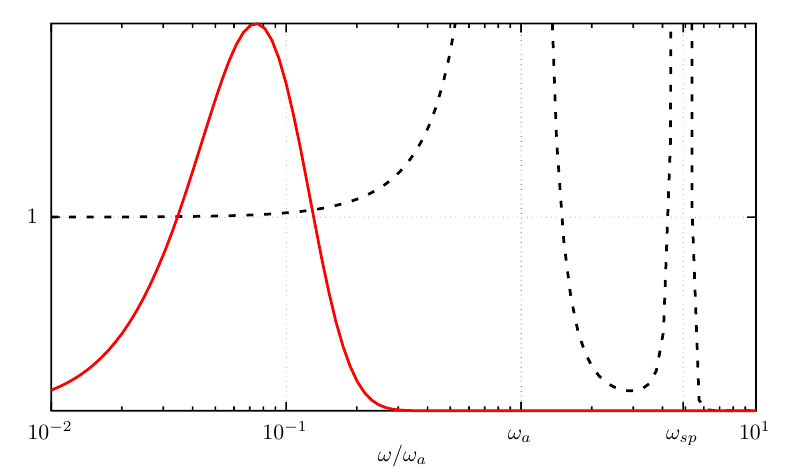}
	\caption{\label{fig:rho_omega}
  Scaled representation of the functions composing the integrand in Eq.\eqref{Eq:ForceDipoleFilter} for the thermal viscosity.
  Dashed black line:
  Spectral density of the atom+field system $ \eta(\omega)$ [see Eq.~\eqref{Eq:RhoEM}] normalized to its low-frequency  behavior [see Eq.~\eqref{Eq:rhoEMSurfaceLow}].
  We again consider the planar setup in the near field and mark the surface resonance $\omega_{\mathrm{sp}}$ as well as the single atomic resonance $\omega_a$.
 Full red line:
  $T\partial_T\varrho(\omega)$ evaluated at room temperature divided by half its maximum value.
  Depending on the chosen value of $T$, the peak of $T\partial_T\varrho(\omega)$ can pronounce qualitatively different parts of the spectral density [see main text and Eq.~\eqref{Eq:ForceDipoleFilter}].
  Setup and parameters are chosen as in Fig.~\ref{fig:z_T} and we use $z_a=1$ nm.
 }
\end{figure}
%%%%%%%%%%%%%%%%%%%5
%\end{comment}

To see this more clearly, let us once again consider the case of a single-resonance polarizability for an atom moving in the near-field of a flat surface. At the leading order atom-field-coupling, Eq.~\eqref{Eq:RhoEM} takes the form \cite{tomas95,intravaia2016a,intravaia2019}
\begin{align}
\label{Eq:RhoEMExample}
& \eta(\omega)
  \sim
  \frac{9c^3}{4\pi\epsilon_0^2\omega^4}
  |\alpha_{\mathrm{B}}(\omega)|^2
\frac{\mathrm{Im}\left[r(\omega)\right]^2}{(2z_a)^8}.
\end{align}
For complementarity, in the previous expression, we considered only the case of sufficiently small distance between atom and surface, where the homogeneous part of the Green tensor can be safely neglected and only the scattering part, which is connected to the interface, has to be considered.
Also, for simplicity, we restricted our discussion to spatially local material models \cite{reiche2017}.
Equation \eqref{Eq:RhoEMExample} highlights that $ \eta(\omega)$ contains resonances related to the particle ($\sim\alpha_{\mathrm{B}}$) for $\omega\sim \omega_a$ and, if they exist, to the surface ($\sim \mathrm{Im}\left[r\right]$). Examples of the latter are surface-plasmon- and surface-phonon-polariton resonances \cite{intravaia2016a} appearing, respectively, in the case of metallic or dielectric surfaces.
At frequencies smaller than any of these resonances, $ \eta(\omega)$ is still finite due the dissipation in the material, i.e.
\begin{align}
\label{Eq:rhoEMSurfaceLow}
& \eta(\omega)
  \sim
  \frac{9}{\pi}
  \frac{\alpha_0^2}{(2z_a)^8}
  \frac{\varrho^2}{\omega^2}.
\end{align}
In Fig.~\ref{fig:rho_omega} we depict both the functions in the integrand of Eq.~\eqref{Eq:ForceDipoleFilter} in the case of a $^{87}$Rb atom moving above a gold surface (their normalization is different in order to better visualize the overlap region).
Depending on the temperature, the function $T\partial_T\varrho(\omega)$ selects the resonant region of $\eta(\omega)$ or the dissipative low-frequency regime. This gives rise to the resonant or the non-resonant form of frictional interaction, respectively.
For comparison, if we take room temperature as a reference, the thermal distribution has a maximum around 7\%
%5.6 \% {\color{red}(Check: Did you calculate this value using $\omega_{a}=1.3$  eV?)}
the value of the resonance frequency of the $^{87}$Rb  D line \cite{steck}, substantially overlapping with the low-frequency part of $ \eta(\omega)$. For this specific case, one would need a temperature at least twenty times larger for the maximum of $T\partial_T\varrho(\omega)$ to reach the value of the atomic transition $\omega_{a}\sim \fran{1.3}$ eV or that of the surface plasmon resonance $\omega_{\rm sp}\sim 6.4$ eV. In this case, the frictional viscosity $\mu_T$ would be characterized by its resonant contribution.
%to the interaction dominate and hence $\mu_T^{\mathrm{vac}}$ will prevail as the dominating contribution to the thermal viscosity. This regime corresponds to the right area in Fig.~\ref{fig:z_T}.

%
% CONCLUSION
%

\vspace{0.4cm}
\section{Conclusion}
\label{sec:conclusion}

An atom driven at constant speed $v$ in the near a complex electromagnetic environment comprised by several translationally invariant objects at temperature $T$ experiences a force that opposes its motion as if it were moving in a viscous medium. The quantum and thermal fluctuations of the electromagnetic field define a privileged reference frame with respect to which the atom tends to be at rest.
Depending on the temperature, the optical response and the geometry of its surroundings as well as  the spin angular momentum of the radiation, the perceived viscosity is characterized by different physical phenomena.
One can distinguish three distinct regimes. (i) At sufficiently large distance $z_a$ from any material interface, the viscosity is dominated by the interaction with the thermal vacuum field. That particular limit recovers what is known in the literature as black-body friction \cite{einstein10a,mkrtchian2003,lach12a}. Black-body friction scales linearly in velocity $v$ and the corresponding viscosity depends on the temperature and the parameters characterizing the atom and the electromagnetic radiation in vacuum only.
(ii) The closer the particle moves to an interface, the more the interaction is affected by the change in the electromagnetic density of states due to the boundary conditions induced by the material-induced inhomogeneity of space.
At sufficiently low velocities, large separation or high temperatures ($k_{\mathrm{B}}T\gg\hbar v/\lambda_a$), the frictional interactions is still mainly thermal.
In addition to the parameters determining the drag in vacuum, the viscosity now also depends on the optical properties and geometry of the materials characterizing the electromagnetic environment and on the position of the atom.
At the leading order the frictional force scales linearly in the velocity and therefore in this limit the viscosity does not depend on $v$.
Finally, (iii) in the limit of sufficiently low temperatures or, comparably, short distances or high velocities ($k_{\mathrm{B}}T\ll\hbar v/\lambda_a$), electromagnetic quantum fluctuations become the main source of the interaction, recovering the behavior of quantum friction \cite{volokitin2007,intravaia11a,dedkov2017}: The quantum drag force is $\propto v^{3}$, corresponding to a viscosity that scales as $v^{2}$.

The description presented here unifies and extends the results that are presently available in the the literature by self-consistently and non-perturbatively including both macroscopic bodies and a spin-dependent interaction between atom and the material-modified field.
Similarly to what has already been done for zero temperature \cite{intravaia2019}, our formalism enabled us to identify a contribution to the frictional processes that involve the exchange of angular momentum between the confined field in the vicinity of the material interface and the particle.
At nonzero temperature the transfer of angular momentum turns out to reduce the corresponding viscosity by a factor one half (at zero temperature the reduction was $5/7$). We also highlighted the characteristic scales which describe the transition among the different regimes [(i), (ii), and (iii)] of the frictional interaction, pointing out that the quantum properties of the drag force become more relevant for particles with a high kinetic energy.
We explained this result using the anomalous-Doppler-effect through which virtual excitations (quantum fluctuations) can receive enough energy to the cost of the atom's kinetic energy to turn into real ones  \cite{ginzburg1960,nezlin1976,intravaia2015b,intravaia2019}.

From the experimental point of view, our results will be of relevance for high-precision experiments with atoms, such as in the design of frequency references \cite{ludlow15} or in the interpretation atom interferometric measurements \cite{abend20a}, where black-body friction, e.g., can be connected to Stark shifts and modified atomic lifetimes \cite{farley81,lach12}.
Especially with the advent of miniaturization efforts in order to create portable devices of these emerging quantum technologies \cite{schuldt21,bongs19}, the impact of the instrument's boundaries starts to matter
and deviations from the simple Planck spectrum can become important \cite{kalensher84,garciagarcia08}.
For instance, if we take once again the simplest case of a rubidium atom moving in front of a single gold surface (described via the Drude model) at room temperature, the drag connected to the interaction with the surface starts to exceed the black-body vacuum drag below separations of some tenths of micrometers.
The results and the description outlined in the present manuscript provide the tools that can allow to better understand and possibly engineer the vacuum viscosity in the presence of (shape-)optimized materials via advanced numerical schemes.

%%%%%%%%%%%%%%%%%%%%%%%%%%%%%%%%%%%%%%%%%%%%%%%%%%%%%%%%%%%%%%%%%%%%%%
%
% ACKNOWLEDGMENTS
%
\begin{acknowledgements}
MO and DR have contributed equally to this work.
We thank Bettina Beverungen and Dan-Nha Hyunh for helpful discussions.
DR thanks Simon Kanthak, Julien Kluge, Markus Krutzik, Vladimir Schkolnik and Aaron Strangfeld for insights on atomic spectroscopy and is grateful to Bei-Lok Hu for interesting discussions about the viscosity of vacuum.
\end{acknowledgements}

%%%%%%%%%%%%%%%Appendix%%%%%%%%%%%%%%%%%%%%%%%%%%

%
% APPENDICES
% ----------
%

\appendix
 \section{Mathematical details \\on derivation of frictional force}
 \label{App:DerForceGen}
Here we provide additional information about the derivation of some of the expressions presented in the manuscript.
We start by considering Eq.~\eqref{Eq:ForceDipoleLate}. Following Ref. \cite{intravaia2016a}, the Lorentz force acting on the moving dipole can be written as
\begin{align}
\label{Lorentz}
F(t)=
\lim_{\vv r \to \vv r_a(t)}
2\mathrm{Re}\;
\langle
\hat{\vv d}(t)\cdot
\partial_x
\hat{\vv{E}}^{\oplus}(\vv r, t)
\rangle,
\end{align}
where $\hat{\vv{E}}^{\oplus}(\vv r, t)$ is the part of the total electric field operator related only with an integration over positive frequencies $\omega\geq 0$.
As described in Sec. \ref{sec:noncontact}, the electric field operator decomposes into two components, i.e.
$
  \hat{\mathbf{E}}(\mathbf{r},t)
  =
  \hat{\mathbf{E}}_0(\mathbf{r},t)
  +
  \hat{\mathbf{E}}_\mathrm{ind}(\mathbf{r},t)
$,
and each of them can be split in an integration over positive and negative frequencies $\hat{\mathbf{E}}^{\oplus}$ and $\hat{\mathbf{E}}^{\ominus}$, respectively.
For $\hat{\mathbf{E}}_0$ we can write
\begin{align}
\hat{\mathbf{E}}_0^{\oplus}(\mathbf{r}_{a}(t),t)
  &=
  \int_0^{\infty}\frac{\mathrm{d}\omega}{2\pi}
  \int\frac{\mathrm{d}q}{2\pi}~
  \hat{\mathbf{E}}_0(q,\mathbf{R}_a,\omega)
  e^{\ii(q x_a(t)-\omega t)}
\end{align}
and from Eq.~\eqref{eq:efield} we can obtain the corresponding relation for the field induced by the dipole, $\hat{\mathbf{E}}_\mathrm{ind}^{\oplus}(\mathbf{r},t)$.
Accordingly, the force (formally) in Eq.~\eqref{Lorentz} also decomposes into two parts.
In the steady-state ($t\to \infty$), for the contribution to the force connected to $\hat{\mathbf{E}}_0^{\oplus}(\mathbf{r},t)$, we hence obtain

\begin{align}
&
\lim_{\vv r \to \vv r_a(t)}
2\mathrm{Re}\;
\langle
\hat{\vv d}(t)\cdot
\partial_x
\hat{\vv{E}}^{\oplus}_0(\vv r, t)
\rangle
\\\nonumber
  &=
  2
  \mathrm{Re}\;
  \int\frac{\mathrm{d}\tilde{\omega}}{2\pi}
  \int_0^{\infty}\frac{\mathrm{d}\omega}{2\pi}
  \int\frac{\mathrm{d}q}{2\pi}~
  (\ii q)
  \\\nonumber
  &\quad\times
  \langle
  \hat{\vv d}(\tilde{\omega})\cdot
  \hat{\mathbf{E}}_0(q,\mathbf{R}_a,\omega)
  \rangle
  e^{\ii(q x_a(t)-[\omega+\tilde{\omega}] t)}
  e^{\ii \tilde{q}x_0}
  \\\nonumber
    &=
    2
    \mathrm{Re}\;
    \int\frac{\mathrm{d}\tilde{\omega}}{2\pi}
    \int_0^{\infty}\frac{\mathrm{d}\omega}{2\pi}
    \int\frac{\mathrm{d}q}{2\pi}
    \int\frac{\mathrm{d}\tilde{q}}{2\pi}~
    (\ii q)
    e^{-\ii(\omega+\tilde{\omega}-q v)t}
    \\\nonumber
    &\quad\times
    \mathrm{Tr}
    \left[
    \underline{\alpha}_v(\tilde{\omega})
    \langle
    \hat{\mathbf{E}}_0(\tilde{q},\mathbf{R}_a,\tilde{\omega}+\tilde{q}v)
    \hat{\mathbf{E}}^{\intercal}_0(q,\mathbf{R}_a,\omega)
    \rangle
    \right]
    e^{\ii (q+\tilde{q})x_0}
    .
\end{align}
In the previous expression we used that in the steady state the atom's trajectory is given $x_a(t)\sim x_0+ v t$, where $x_{0}$ is the position where system becomes stationary. 
We also use that the scalar product of two vectors $\mathbf{a}$ and $\mathbf{b}$ is identical to the trace of the dyadic $\mathbf{a}\mathbf{b}^{\intercal}$, i.e., $\mathbf{a}\cdot\mathbf{b}=\mathrm{Tr}[\mathbf{a}\mathbf{b}^{\intercal}]$, and that the steady state solution for the dynamics of the dipole operator $\hat{\mathbf{d}}(t)$ is given by the stationary solution of Eq.~\eqref{eq:deom}. In frequency domain [see Eq.~\eqref{Eq:FDTNEq}] its expression is
\begin{align}
\hat{\mathbf{d}}(\tilde{\omega})
  &=
  \underline{\alpha}_v(\tilde{\omega})
  \int\frac{\mathrm{d}\tilde{q}}{2\pi}~
  \hat{\mathbf{E}}_0(\tilde{q},\mathbf{R}_a,\tilde{\omega}+\tilde{q}v)
  e^{\ii \tilde{q} x_0}.
\end{align}
Since $\hat{\mathbf{E}}_0$ describes the field withouts the atom, the dyadic $ \langle
    \hat{\mathbf{E}}_0(\tilde{q},\mathbf{R}_a,\tilde{\omega}+\tilde{q}v)
    \hat{\mathbf{E}}^{\intercal}_0(q,\mathbf{R}_a,\omega)
    \rangle$ can be evaluated using the fluctuation-dissipation relation in Eq.~\eqref{eq:flucdisE0}. Finally, we obtain
\begin{align}
&
\lim_{\vv r \to \vv r_a(t)}
2\mathrm{Re}\;
\langle
\hat{\vv d}(t)\cdot
\partial_x
\hat{\vv{E}}^{\oplus}_0(\vv r, t)
\rangle
\\\nonumber
    &=
    4\hbar
    \mathrm{Re}\;
    \int_0^{\infty}\frac{\mathrm{d}\omega}{2\pi}
    \int\frac{\mathrm{d}q}{2\pi}
    \ii q\,
    n(\omega)
    \mathrm{Tr}
    \left[
    \underline{\alpha}_v(-\omega_q^-)
    \underline{G}_{\Im}^{\intercal}(q,\mathbf{R}_a,\omega)
    \right]
    ,
\end{align}
where we used that
$-\underline{G}_{\Im}^{\intercal}(q,\mathbf{R}_a,\omega)=\underline{G}_{\Im}(-q,\mathbf{R}_a,-\omega)$, $n(-\omega)=-[1+n(\omega)]$.
If we now evaluate the real part of the expression, we arrive at

\begin{align}
&
\lim_{\vv r \to \vv r_a(t)}
2\mathrm{Re}\;
\langle
\hat{\vv d}(t)\cdot
\partial_x
\hat{\vv{E}}^{\oplus}_0(\vv r, t)
\rangle
\\\nonumber
    &=
    -4\hbar
    \int_0^{\infty}\frac{\mathrm{d}\omega}{2\pi}
    \int\frac{\mathrm{d}q}{2\pi}
    ~q
    n(\omega)
    \\\nonumber
    &\times
    \mathrm{Tr}
    \left[
    \frac{
    \underline{\alpha}_v(-\omega_q^-)
    \underline{G}_{\Im}^{\intercal}(q,\mathbf{R}_a,\omega)
    -
    \underline{\alpha}_v^*(-\omega_q^-)
    \underline{G}_{\Im}^{\dagger}(q,\mathbf{R}_a,\omega)
    }{2\ii}
    \right]
    \\\nonumber
        &=
        -2\frac{\hbar}{\pi}
        \int_0^{\infty}\mathrm{d}\omega
        \int\frac{\mathrm{d}q}{2\pi}
        ~q
        n(\omega)
        \mathrm{Tr}
        \left[
        \underline{\alpha}_{v,\Im}(-\omega_q^-)
        \underline{G}_{\Im}^{\intercal}(q,\mathbf{R}_a,\omega)
        \right]
    ,
\end{align}
where we used in the second line that $\underline{G}_{\Im}(q,\mathbf{R}_a,\omega)=\underline{G}_{\Im}^{\dagger}(q,\mathbf{R}_a,\omega)$ and that the trace of a matrix is identical to the trace of its transpose.
The previous result gives the first term in Eq.~\eqref{Eq:ForceDipoleLate}.
The second term of Eq.~\eqref{Eq:ForceDipoleLate} as been derived in earlier work \cite{intravaia2016a}. Combining the two, we arrive at Eq.~\eqref{Eq:ForceDipoleLate}.

Let us consider now our result in Eq.~\eqref{Eq:ForceDipoleLate2}.
Starting from Eq.~\eqref{Eq:ForceDipoleLate}, we first substitute $\omega\to\omega+qv=\omega_q^+$, i.e.
\begin{align}
\label{Eq:Aux0}
F&
  =
  -
  2
  \int\frac{\mathrm{d}q}{2\pi}
  \int_{-qv}^{\infty}\mathrm{d}\omega
  q
  \\\nonumber
  &~\times
  \mathrm{Tr}
  \left[
  \left\{
  \frac{\hbar}{\pi}
  n(\omega_q^+)
  \underline{\alpha}_{v,\Im}(-\omega)
  +
  \underline{S}_{v}(-\omega)
  \right\}
  \underline{G}_{\Im}^{\intercal}(q,\mathbf{R}_a,\omega_q^+)
  \right].
\end{align}
Second, we employ the identities $\underline{\alpha}_{v,\Im}(-\omega)=-\underline{\alpha}_{v,\Im}^{\intercal}(\omega)$ and $\underline{S}_v(-\omega)=\underline{S}_v^{\intercal}(\omega)-(\hbar/\pi)\underline{\alpha}_{\Im}^{\intercal}(\omega)$ to obtain
\begin{align}
\label{Eq:Aux1}
  &F
  =
  2
  \int\frac{\mathrm{d}q}{2\pi}
  \int_{-qv}^{\infty}\mathrm{d}\omega
  ~q
  \\\nonumber
  &\times
  \mathrm{Tr}
  \left[
  \left\{
  \frac{\hbar}{\pi}
  [1+
  n(\omega_q^+)
  ]
  \underline{\alpha}_{v,\Im}^{\intercal}(\omega)
  -
  \underline{S}_{v}^{\intercal}(\omega)
  \right\}
  \underline{G}_{\Im}^{\intercal}(q,\mathbf{R}_a,\omega_q^+)
  \right].
\end{align}
The properties of the trace operation allow to remove the ``$\intercal$'' superscript in the previous expression.
The only difference with respect to Eq.~\eqref{Eq:ForceDipoleLate2} is that the integral over the frequencies which runs from $-qv$ instead of zero.
Interestingly, however, upon substituting $\omega\to -\omega$ and $q\to -q$, we have that
\begin{align}
&\int\frac{\mathrm{d}q}{2\pi}
\int_{-qv}^{0}\mathrm{d}\omega
~q
\\\nonumber
&\times
\mathrm{Tr}
\left[
\left\{
\frac{\hbar}{\pi}
[1+
n(\omega_q^+)
]
\underline{\alpha}_{v,\Im}(\omega)
-
\underline{S}_{v}(\omega)
\right\}
\underline{G}_{\Im}(q,\mathbf{R}_a,\omega_q^+)
\right]
\\\nonumber
&=
\int\frac{\mathrm{d}q}{2\pi}
\int_{-qv}^{0}\mathrm{d}\omega
~q
\\\nonumber
&\times
\mathrm{Tr}
\left[
\left\{
\frac{\hbar}{\pi}
n(\omega_q^+)
\underline{\alpha}_{v,\Im}(-\omega)
+
\underline{S}_{v}(-\omega)
\right\}
\underline{G}_{\Im}^{\intercal}(q,\mathbf{R}_a,\omega_q^+)
\right]
\\\nonumber
&=
-
\int\frac{\mathrm{d}q}{2\pi}
\int_{-qv}^{0}\mathrm{d}\omega
~q
\\\nonumber
&\times
\mathrm{Tr}
\left[
\left\{
\frac{\hbar}{\pi}
[
1+
n(\omega_q^+)
]
\underline{\alpha}_{v,\Im}(\omega)
-
\underline{S}_{v}(\omega)
\right\}
\underline{G}_{\Im}(q,\mathbf{R}_a,\omega_q^+)
\right].
\end{align}
Here, we used again the identities for the polarizability and the power spectrum given after Eq.~\eqref{Eq:Aux0} as well as those for the Bose number and the Green tensor given after Eq.~\eqref{Eq:ForceDipoleLate2}. The above result indicates that this contribution is zero and that we can replace the frequency integration range in Eq.~\eqref{Eq:Aux1} from $[-qv,\;{\infty})$ to $[0,\;{\infty})$, recovering Eq.~\eqref{Eq:ForceDipoleLate2}.

It is also interesting to show that the previous results can be obtained using the symmetric ordering of operators instead of the normal ordering as done in the main text. Our starting point is the Lorentz force given by Eq.~\eqref{Eq:ForceDipole} of the main text, written in the symmetrized form

\begin{align}
\label{symmetricLorentz}
F(t)
  &=
  \lim_{\mathbf{r}\to\mathbf{r}_a}
  \langle
  \hat{\mathbf{d}}(t)\cdot\partial_x\hat{\mathbf{E}}(\mathbf{r},t)
  \rangle_{\mathrm{sym}},
\end{align}
where the symmetric average is defined as

\begin{align}
\langle \hat{\mathbf{A}}\cdot\hat{\mathbf{B}}\rangle_{\mathrm{sym}}
&\equiv \frac{\langle \hat{\mathbf{A}}\cdot\hat{\mathbf{B}}+ \hat{\mathbf{B}}\cdot\hat{\mathbf{A}}\rangle}{2}
= \frac{\langle \hat{A}_{i}\hat{B}_{i}+ \hat{B}_{i}\hat{A}_{i}\rangle}{2}
\end{align}
(in the last line we implicitly summed over the repeated indices).
While the operator ordering is irrelevant in Eq.~\eqref{symmetricLorentz}, this is no longer
true if we consider again the splitting
$
  \hat{\mathbf{E}}(\mathbf{r},t)
  =
  \hat{\mathbf{E}}_0(\mathbf{r},t)
  +
  \hat{\mathbf{E}}_\mathrm{ind}(\mathbf{r},t)
$ and calculated the corresponding terms independently.

We start with the contribution to the force connected to the vacuum part of the field operator $\hat{\mathbf{E}}_0$ and obtain at late times
\begin{align}
&\lim_{\mathbf{r}\to\mathbf{r}_a(t)}
\langle
\hat{\mathbf{d}}(t)\cdot\partial_x\hat{\mathbf{E}}_0(\mathbf{r},t)
\rangle_{\mathrm{sym}}
  \\\nonumber
  &=
  \lim_{\mathbf{r}\to\mathbf{r}_a}
  \int\frac{\mathrm{d}\omega}{2\pi}
  e^{-\ii\omega t}
  \langle
\hat{\mathbf{d}}(\omega)\cdot\partial_x\hat{\mathbf{E}}_0(\mathbf{r},t)
  \rangle_{\mathrm{sym}}
  \\\nonumber
  &=
  \int\frac{\mathrm{d}\omega}{2\pi}
  \int\frac{\mathrm{d}\tilde{\omega}}{2\pi}
  \int\frac{\mathrm{d}\tilde{q}}{2\pi}~
  (\ii \tilde{q})
  e^{-\ii\omega t}
  e^{\ii(\tilde{q} x_a(t)-\tilde{\omega} t)}
  \\\nonumber
  &\qquad\times
  \langle
  \hat{\mathbf{d}}(\omega)\cdot
  \hat{\mathbf{E}}_0(\tilde{q},\mathbf{R}_a,\tilde{\omega})
  \rangle_{\mathrm{sym}}
  \\\nonumber
  &=
  \int\frac{\mathrm{d}\omega}{2\pi}
  \int\frac{\mathrm{d}\tilde{\omega}}{2\pi}
  \int\frac{\mathrm{d}\tilde{q}}{2\pi}~
  \int\frac{\mathrm{d}q}{2\pi}~
  e^{\ii(\tilde{q}v-\tilde{\omega}-\omega)t}
  e^{\ii (q+\tilde{q}) x_0}
  \\\nonumber
  &\quad\times
  (\ii \tilde{q})
  \mathrm{Tr}\left[\underline{\alpha}_v(\omega)
  \langle
  \hat{\mathbf{E}}_0(q,\mathbf{R}_a,\omega+qv)
  \hat{\mathbf{E}}^{\intercal}_0(\tilde{q},\mathbf{R}_a,\tilde{\omega})
  \rangle_{\mathrm{sym}}\right].
\end{align}
In the case of the dyadic product of vector operators, the symmetric average has to be understood componentwise, i.e. $[\langle \hat{\mathbf{A}} \hat{\mathbf{B}}^{\intercal}\rangle_{\mathrm{sym}}]_{ij}=\langle A_{i}B_{j}+B_{j}A_{i}\rangle/2$. In terms of the matrices resulting from the dyadic product, this can be written as
$\langle \hat{\mathbf{A}} \hat{\mathbf{B}}^{\intercal}\rangle_{\mathrm{sym}}=\langle \hat{\mathbf{A}} \hat{\mathbf{B}}^{\intercal}\rangle+\langle \hat{\mathbf{B}}\hat{\mathbf{A}}^{\intercal} \rangle^{\intercal}/2$.
The symmetric average of the vacuum field can be computed by replacing $2(1+n(\omega))\to \coth[\hbar\omega/(2k_{\mathrm{B}}T)]$ in Eq. \eqref{eq:flucdisE0}.
This yields
\begin{align}
\label{Eq:Aux2}
\lim_{\mathbf{r}\to\mathbf{r}_a(t)}
&
\langle
\hat{\mathbf{d}}(t)\cdot\partial_x\hat{\mathbf{E}}_0(\mathbf{r},t)
\rangle_{\mathrm{sym}}
\\\nonumber
  &=
  -\hbar
  \int\frac{\mathrm{d}\omega}{2\pi}
  \int\frac{\mathrm{d}q}{2\pi}~
  (\ii q)
  \coth\left(\frac{\hbar\omega_q^+}{2k_{\mathrm{B}}T}\right)
  \\\nonumber
  &\quad\times
  \mathrm{Tr}
  \left[
  \underline{\alpha}_v(\omega)
  \underline{G}_{\Im}(q,\mathbf{R}_a,\omega_q^+)
  \right]
  \\\nonumber
    &=
    -\hbar
    \int_0^{\infty}\frac{\mathrm{d}\omega}{2\pi}
    \int\frac{\mathrm{d}q}{2\pi}~
    (\ii q)
    \coth\left(\frac{\hbar\omega_q^+}{2k_{\mathrm{B}}T}\right)
    \\\nonumber
    &\quad\times
    \left\{
    \mathrm{Tr}
    \left[
    \underline{\alpha}_v(\omega)
    -
    \underline{\alpha}_v^{\dagger}(\omega)
    \right\}
    \underline{G}_{\Im}(q,\mathbf{R}_a,\omega_q^+)
    \right]
    \\\nonumber
      &=
      \frac{\hbar}{\pi}
      \int_0^{\infty}\mathrm{d}\omega
      \int\frac{\mathrm{d}q}{2\pi}~
      q
      \coth\left(\frac{\hbar\omega_q^+}{2k_{\mathrm{B}}T}\right)
      \\\nonumber
      &\qquad\times
      \mathrm{Tr}
      \left[
      \underline{\alpha}_{v,\Im}(\omega)
      \underline{G}_{\Im}(q,\mathbf{R}_a,\omega_q^+)
      \right].
\end{align}

For the part of the force connected to $\hat{\mathbf{E}}_{\mathrm{ind}}$, we employ Eq. \eqref{eq:efield} and use that  at late times we can write
\begin{align}
&\hat{\mathbf{E}}_\mathrm{ind}(\mathbf{r},t)
=
\int\limits_{-\infty}^{t-t_0} \mathrm{d}\tau\,
\underline{G}(\mathbf{r}(t),\mathbf{r}_a(t-\tau),\tau)\,
\hat{\mathbf{d}}(t-\tau)
\\\nonumber
&\stackrel{t,-t_{0}\to \infty}{=}
\int \mathrm{d}\tau\,
\int\frac{\mathrm{d}\omega}{2\pi}
\int\frac{\mathrm{d}\omega'}{2\pi}
\int\frac{\mathrm{d}q}{2\pi}
\underline{G}(q,\mathbf{R}_a,\tilde{\omega})\,
\hat{\mathbf{d}}(\omega)
\\\nonumber
&\qquad
\times
e^{-\ii (\tilde{\omega}-\omega)\tau-\ii\omega t}
e^{\ii q [x-x_a(t-\tau)]}.
\end{align}
where we could extend the convolution in the first line to $\tau\to-\infty$ since the Green tensor is a causal function.
Upon inserting the steady state the atom's trajectory  $x_a(t)\sim x_0+ v t$ into the symmetrized expression of the force, we obtain
\begin{align}
\label{Eq:Aux3}
&\lim_{\mathbf{r}\to\mathbf{r}_a(t)}
\langle
\hat{\mathbf{d}}(t)\cdot\partial_x\hat{\mathbf{E}}_{\mathrm{ind}}(\mathbf{r},t)
\rangle_{\mathrm{sym}}
\\\nonumber
  &\sim
  \int \mathrm{d}\tau\,
  \int\frac{\mathrm{d}\omega}{2\pi}
  \int\frac{\mathrm{d}\tilde{\omega}}{2\pi}
  \int\frac{\mathrm{d}\nu}{2\pi}
  \int\frac{\mathrm{d}q}{2\pi}
  (\ii q)
  \\\nonumber
  &\times
  \langle
  \hat{\mathbf{d}}(\nu)\cdot
  \underline{G}(q,\mathbf{R}_a,\tilde{\omega})\,
  \hat{\mathbf{d}}(\omega)
  \rangle_{\mathrm{sym}}
  e^{-\ii (\tilde{\omega}-\omega-qv)\tau-\ii(\omega+\nu)t}
  \\\nonumber
  &=
  \hbar
  \int\frac{\mathrm{d}\omega}{2\pi}
  \int\frac{\mathrm{d}q}{2\pi}
  \int\frac{\mathrm{d}\tilde{q}}{2\pi}
  (\ii q)
\coth\left(\frac{\hbar\omega_{\tilde{q}}^{+}}{2k_{\mathrm{B}}T}\right)
  \\\nonumber
  &\times
  \mathrm{Tr}
  \left[
  \underline{\alpha}_v(\omega)
\underline{G}_{\Im}(\tilde{q},\mathbf{R}_a,\omega_{\tilde{q}}^{+})
  \underline{\alpha}_v^{\dagger}(\omega)
  \underline{G}(q,\mathbf{R}_a,\omega_q^{+})
  \right]
  \\\nonumber
  &=
  \hbar
  \int_0^{\infty}\frac{\mathrm{d}\omega}{2\pi}
  \int\frac{\mathrm{d}q}{2\pi}
  \int\frac{\mathrm{d}\tilde{q}}{2\pi}
  (\ii q)
\coth\left(\frac{\hbar\omega_{\tilde{q}}^{+}}{2k_{\mathrm{B}}T}\right)
  \\\nonumber
  &\times
  \left\{
  \mathrm{Tr}
  \left[
  \underline{\alpha}_v(\omega)
\underline{G}_{\Im}(\tilde{q},\mathbf{R}_a,\omega_{\tilde{q}}^{+})
  \underline{\alpha}_v^{\dagger}(\omega)
  \underline{G}(q,\mathbf{R}_a,\omega_q^{+})
  \right]
  \right.
  \\\nonumber
  &
  \quad-
  \left.
  \mathrm{Tr}
  \left[
  \underline{\alpha}_v^{*}(\omega)
\underline{G}_{\Im}^{\intercal}(\tilde{q},\mathbf{R}_a,\omega_{\tilde{q}}^{+})
  \underline{\alpha}_v^{\intercal}(\omega)
  \underline{G}^*(q,\mathbf{R}_a,\omega_q^{+})
  \right]
  \right\}
  \\\nonumber
  &=
  -\frac{\hbar}{\pi}
  \int_0^{\infty}\mathrm{d}\omega
  \int\frac{\mathrm{d}q}{2\pi}
  \int\frac{\mathrm{d}\tilde{q}}{2\pi}
  ~q
\coth\left(\frac{\hbar\omega_{\tilde{q}}^{+}}{2k_{\mathrm{B}}T}\right)
  \\\nonumber
  &\qquad\times
  \mathrm{Tr}
  \left[
  \underline{\alpha}_v(\omega)
\underline{G}_{\Im}(\tilde{q},\mathbf{R}_a,\omega_{\tilde{q}}^{+})
  \underline{\alpha}_v^{\dagger}(\omega)
  \underline{G}_{\Im}(q,\mathbf{R}_a,\omega_q^{+})
  \right]
\end{align}
where, together with the properties of the trace operation, we used that $\underline{G}(-q,\mathbf{R}_a,-\omega)=\underline{G}^*(q,\mathbf{R}_a,\omega)$ and that, similarly as the electromagnetic field, the symmetric dipole correlator can be obtained by replacing in Eq.~\eqref{Eqs:SpecPol} $n(\omega_q^+) +1\to  \coth[\hbar\omega_q^+/(2k_{\mathrm{B}}T)]/2$.
Adding Eqs. \eqref{Eq:Aux2} and \eqref{Eq:Aux3} we arrive at Eq.~\eqref{Eq:ForceDipoleLate2.5}, which is equivalent to the expressions in Eqs.~\eqref{Eq:ForceDipoleLate} and \eqref{Eq:ForceDipoleLate2} obtained using a different approach and ordering scheme.
%
%

%
% BIBLIOGRAPHY
%
%\bibliography{./bib/lib}
%\bibliographystyle{./bib/bibstyle/prstytitlenew}

\end{document}